\documentclass[
showpacs,
preprintnumbers,
reprint,
amsmath,amssymb,
10pt
,letterpaper
,aip,jcp, 
citeautoscript
]
{revtex4-1}

\usepackage{graphicx}
\usepackage{dcolumn}
\usepackage{bm}

\usepackage[english]{babel} 
\usepackage{xcolor}

 \newcommand{\revi}[1]{{#1}}

\renewcommand{\vec}[1]{{\bm{#1}}} 
\newcommand{\uvec}[1]{{\mathbf{\hat{#1}}}} 

\newcommand{\vect}[1]{{\mathbf{#1}}} 
\newcommand{\vgr}[1]{{\bm{#1}}} 

\newcommand{\erf}{\mathrm{erf}}

\newcommand{\rh}[1]{\rho^{(#1)}}

\newcommand{\curr}[1]{\vec{\mathcal{J}}_{\!\!#1}} 
\newcommand{\currk}[1]{\vec{\mathcal{K}}_{\!#1}} 

\begin{document}

\title{
Multi-species dynamical density functional theory for microswimmers: 
derivation, orientational ordering, trapping potentials, and shear cells
}

\author{Christian Hoell}
\email{christian.hoell@uni-duesseldorf.de}

\author{Hartmut L\"owen}

\author{Andreas M.~Menzel}
\email{menzel@hhu.de}
\affiliation{Institut f\"ur Theoretische Physik II: Weiche Materie,
Heinrich-Heine-Universit\"at D\"usseldorf,
Universit\"atsstra\ss{}e~1,
D-40225 D\"usseldorf, Germany.}

\date{\today}

\begin{abstract}
Microswimmers typically operate in complex environments. In biological systems, often diverse species are simultaneously present and interact with each other. Here, we derive a (time-dependent) particle-scale statistical description, namely a dynamical density functional theory, for such multi-species systems, extending existing works on one-component microswimmer suspensions. In particular, our theory incorporates the effect of external potentials, but also steric and hydrodynamic interactions between swimmers. For the latter, a previously introduced force-dipole-based minimal (pusher or puller) microswimmer model is used. As a limiting case of our theory, mixtures of hydrodynamically interacting active and passive particles are captured as well. After deriving the theory, we apply it to different planar swimmer configurations. First, these are binary pusher--puller mixtures in external traps. In the considered situations, we find that the majority species imposes its behavior on the minority species. Second, for unconfined binary pusher--puller mixtures, the linear stability of an orientationally disordered state against the emergence of global polar orientational order (and thus emergent collective motion) is tested analytically. Our statistical approach predicts, qualitatively in line with previous particle-based computer simulations, a threshold for the fraction of pullers and for their propulsion strength that lets overall collective motion arise. Third, we let driven passive colloidal particles form the boundaries of a shear cell, with confined active microswimmers on their inside. Driving the passive particles then effectively imposes shear flows, which persistently acts on the inside microswimmers. Their resulting behavior reminds of the one of circle swimmers, though with varying swimming radii.
\end{abstract}

\maketitle

\section{Introduction}

From a fundamental point of view, the study of active microswimmers \cite{purcell1977life, lauga2009hydrodynamics, elgeti2015physics,menzel2015tuned,zottl2016emergent, bechinger2016active} 
--- i.e., micron-sized self-propelling particles suspended in a fluid --- 
is interesting already because of the inherent non-equilibrium nature of self-propelling particles.\cite{ramaswamy2010mechanics, cates2012diffusive,marchetti2013hydrodynamics, menzel2016way} 
Unusual collective behavior arises from this feature, e.g., motility-induced phase separation (MIPS)\cite{cates2013active,buttinoni2013dynamical,bialke2013microscopic,bialke2015active,cates2015motility,solon2018generalized_NJP,digregorio2018full}
and laning.\cite{wensink2012emergent,menzel2012soft,menzel2013unidirectional,kogler2015lane,romanczuk2016emergent,menzel2016way}
Moreover, on the applied side, natural biological microswimmers  \cite{purcell1977life,kessler1985hydrodynamic,eisenbach2006sperm,engstler2007hydrodynamic,berg2008coli,polin2009chlamydomonas,durham2009disruption, mussler2013effective,goldstein2015green}
occur in almost all locations on Earth, including the human body,
and artificial microswimmers \cite{paxton2004catalytic,howse2007self,buttinoni2012active,walther2013janus,samin2015self,moran2017phoretic}
may in the near future be used in medical and technical applications on the microscale,
e.g.,  for precise drug delivery,\cite{wang2012nano,patra2013intelligent,abdelmohsen2014micro,ma2015catalytic,demirors2018active} 
for non-invasive surgery,\cite{nelson2010microrobots,xi2013rolled,abdelmohsen2014micro} 
when guiding sperm cells, \cite{magdanz2013development} 
and to power microengines.\cite{di2010bacterial,sokolov2010swimming,maggi2016self}

Both biological and artificial microswimmers typically operate under complex conditions. \cite{bechinger2016active}
For example, the complexity can arise from steric confinement of the swimmers \cite{wensink2008aggregation,elgeti2009self,kaiser2012capture,licata2016diffusion,janssen2017aging,praetorius2018active} or be induced by a complex dispersion medium.\cite{gomez2016dynamics,narinder2018memory,liebchen2018viscotaxis,ferreiro2018long,daddi2018dynamics,puljiz2019memory} 
Here, we consider the complementing case of complexity caused by interactions between different swimmer species, as can occur in a diverse set of situations.

In medical contexts, active multi-species systems (including both active--active and active--passive mixtures)
appear when active agents, e.g., pathogenic bacteria or cargo-delivering microrobots, interact with (similar-sized) human cells.
Furthermore, real-world microorganisms can change between motile and non-motile (i.e., active and passive in our notation) behavior during their life,
with the organization in many-particle biofilms \cite{flemming2016biofilms,vidakovic2018dynamic} and active carpets \cite{berke2008hydrodynamic,mathijssen2018self} as examples for extreme cases.
Also, different mutant lines of the same bacterial species can show different motility properties, see, e.g., motile and non-motile strains of \emph{E.~coli} bacteria.\cite{berg2008coli}
More in general, subgroups of swimmers may be identified, if a strong polydispersity, e.g., of swimming speeds, is present inside a system.
Finally, at least two species of swimmers are necessary to construct ``heteronuclear'' (i.e., composed of different building blocks) microswimmer molecules.\cite{babel2016dynamics,kuchler2016getting,lowen2018active}

Despite these manifold possible applications, studies on mixtures of microswimmers (and active particles in general) are still relatively rare.
The problems regarded thus far include predator--prey dynamics,\cite{mecholsky2010obstacle,sengupta2011chemotactic}
mixtures of active rotors with opposite senses of rotation\cite{nguyen2014emergent,yeo2015collective} (see also the corresponding macroscale equivalent in Ref.~\onlinecite{scholz2018rotating}),
transport of passive V-shaped cargo particles by active rods in the bulk \cite{angelani2010geometrically,kaiser2014transport,kaiser2015mechanisms,kaiser2015motion}
and by circle swimmers in channels \cite{liao2018transport},
depletion interactions between passive particles induced by an active bath,\cite{ray2014casimir,ni2015tunable}
segregation effects in mixtures of Taylor-line swimmers propelling by self-deformation,\cite{agrawal2018self}
mixtures in which the activity is introduced by an effective colored noise,\cite{wittmann2018effective}
mesoscale transport phenomena in multi-species microorganism systems,\cite{ben2016multispecies}
and MIPS-like phase separation in active--passive mixtures.\cite{stenhammar2015activity,kummel2015formation,wysocki2016propagating,wittkowski2017nonequilibrium,alaimo2018microscopic}
Furthermore, collective behavior in mixtures of straight-propelling particles\cite{menzel2012collective,guisandez2017heterogeneity} and
those migrating on circular trajectories\cite{levis2018activity, levis2019simultaneous} has been studied 
assuming Vicsek-type \cite{vicsek1995novel,toner1995long,ihle2011kinetic} effective alignment interactions between the swimmers. 
In addition to that, particle-based computer simulations of binary mixtures of microswimmers with different types of propulsion mechanisms,
subject to mutual hydrodynamic interactions, have been performed to quantify the effect on the overall collective alignment behavior.\cite{pessot2018}

In the present work, we cover multi-component microswimmer suspensions subject to external potentials. Different species here are mutually interacting,  both via steric interaction potentials and via (far-field) hydrodynamic interactions.
The latter may, following classical statistical mechanics (for passive particles), affect the dynamic behavior but, in general,  not the appearance of static equilibrium systems.
Microswimmer suspensions, however, are inherently out of equilibrium so that even steady states may be significantly altered by hydrodynamic interactions, 
calling for their incorporation in the physical description.
Additionally, interesting phenomena can appear when hydrodynamic effects interplay with, e.g., magnetic interactions.\cite{babel2016dynamics,guzman2016fission}

Generally, supplementing experiments and many-body particle-based simulations with statistical descriptions, e.g., density-field equations,
allows for thorough theoretical analysis.
Ideally, the observed phenomena are explained in this way and new types of behavior are predicted,
leading to a better understanding of the underlying physical effects.
A well-established way of finding such density-field equations in non-equilibrium colloidal systems is dynamical density functional theory (DDFT).\cite{evans1979nature,evans1992density,marconi1999dynamic,marconi2000dynamic,archer2004dynamical,chan2005time,wensink2008aggregation,espanol2009derivation,evans2010density,lowen2010density,wittkowski2011dynamical,menzel2016dynamical,hoell2017dynamical,hoell2018particle}
Accordingly, we successfully derived a DDFT for one-species microswimmer systems and applied it to several example situations in previous works.\cite{menzel2016dynamical,hoell2017dynamical,hoell2018particle}
In other contexts, DDFTs for mixtures of passive colloidal particles have been developed before.\cite{archer2005dynamical,goddard2013multi,hartel2015anisotropic,babel2018impedance, somerville2018density}
Here, we combine these two approaches and explicitly allow for different species of active microswimmers (and~/~or passive particles).
In addition to the applications listed above, such a DDFT might in the future help to find dynamic correlation functions in one-component systems via ``test-particle'' methods.\cite{archer2007dynamics,hopkins2010van,hoell2018particle}
We remark that multi-species DDFT approaches can also be used to describe the dynamics of other kinds of active matter, e.g., the growth of tumors in cell tissues.\cite{alsaedi2018dynamical}

Below, the employed microswimmer model --- introduced in previous works \cite{menzel2016dynamical,hoell2017dynamical,pessot2018,hoell2018particle} --- and its implications for hydrodynamic interactions are overviewed in Sec.~\ref{sec:model}. 
It is then used in Sec.~\ref{sec:ddft} as an input to derive the statistical theory, namely the multi-species dynamical density functional theory for microswimmers.
Subsequently, several applications of the theory are discussed in Sec.~\ref{sec:results}, where we confine ourselves to planar arrangements within three-dimensional fluids.
First, extending the one-component case analyzed previously,\cite{menzel2016dynamical,hoell2017dynamical} we explore binary mixtures of microswimmers in an external trap and find additional steady states resulting from interspecies interactions.
Second, the possibility of emergent overall orientational order due to hydrodynamic interactions in binary mixtures of microswimmers is discussed.
Third, microswimmers confined inside an externally driven ring of passive colloidal particles are investigated. The passive particles induce a shear flow that the enclosed active swimmers are exposed to.
Finally, a short summary and an outlook are given in Sec.~\ref{sec:con}.

\section{Swimmer model and the resulting hydrodynamic interactions}
\label{sec:model}

Before a particle-scale statistical description can be developed in Sec.~\ref{sec:ddft}, 
an appropriate discretized swimmer model must first be defined.
In particular, the hydrodynamic interactions between individual swimmers are specified below.
For this purpose, we briefly review the previously-introduced minimal swimmer model.\cite{menzel2016dynamical,hoell2017dynamical,pessot2018,hoell2018particle}

Since a microswimmer cannot exert a net force on the surrounding liquid,\cite{purcell1977life,ten2015can}
the far-field fluid flow around a swimmer (to lowest order) can typically be described as if it were caused by a force dipole acting on the fluid.
(Exceptions are ``neutral-type'' swimmers with a vanishing time-averaged force-dipole contribution,\cite{najafi2004simple,zargar2009three,daddi2018swimming,daddi2018state} which only feature higher-order multipole terms in the far-field flow caused, e.g., by an effective force quadrupole.)
Here, we explicitly resolve the force dipole by two oppositely-oriented forces of equal magnitude.

Depending on whether the forces push out or pull in the fluid along the axis of self-propulsion,
one distinguishes between \emph{pusher} (extensile) and \emph{puller} (contractile) microswimmers.\cite{underhill2008diffusion, baskaran2009statistical}
Consequently, pushers draw in the fluid from the transverse directions,
while pullers expel it along them. 
Our model can cover both cases, as detailed below.

Low Reynolds numbers --- as are typical for microswimmers \cite{purcell1977life} --- and incompressibility of the fluid are henceforth assumed.
Particularly, this means that the response of the fluid to a force is linear, overdamped, and instantaneous.
In the bulk, the analytically-known Oseen tensor then explicitly connects the effect of a point-like force center to the resulting fluid flow.\cite{Happel_Brenner,Dhont,Kim_Karrila}
For finite-sized spherical particles subject to net forces and torques, the way to find (approximate) expressions for the induced hydrodynamic interactions between them is well-established.\cite{Dhont, Kim_Karrila}

\begin{figure}[t]
\vspace{-30pt}
\includegraphics{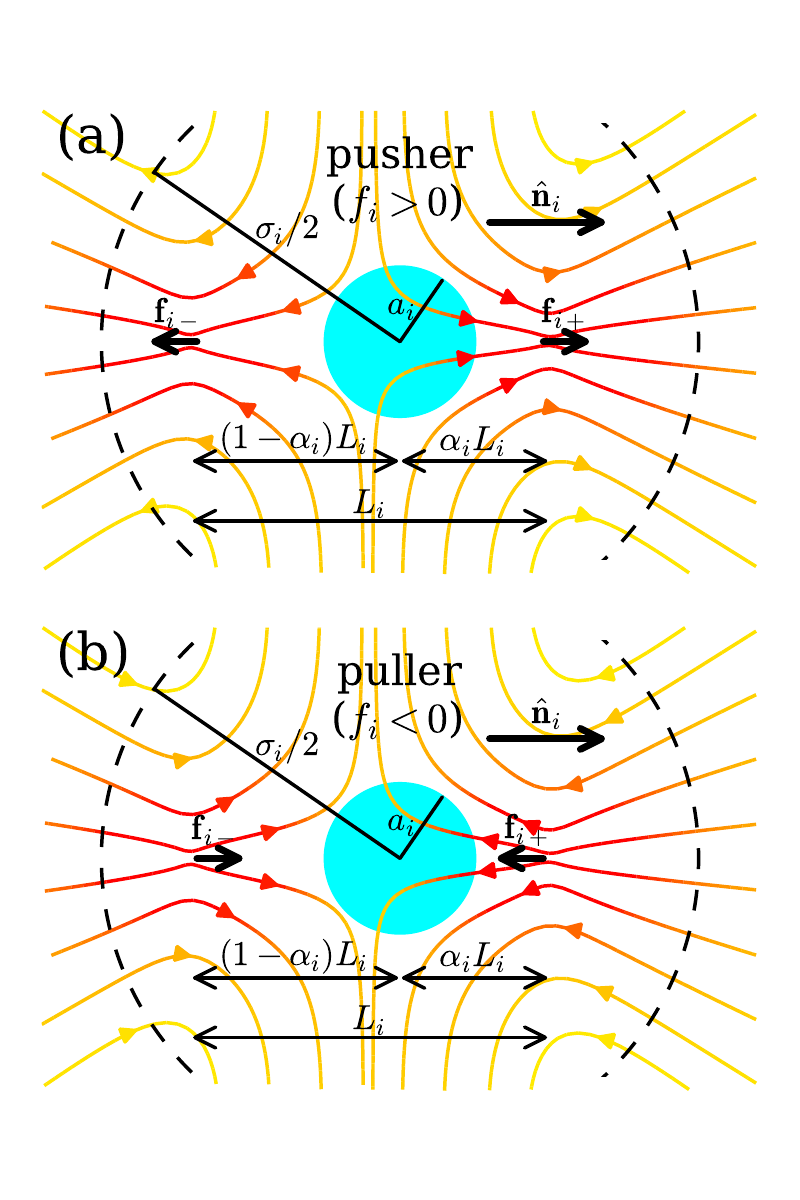}
\vspace{-30pt}
\caption{Force-dipole-based minimal microswimmer model.\cite{menzel2016dynamical}
Around a central sphere of radius~$a_i$, two anti-parallel equal-magnitude forces \mbox{$\vect{f}_{i\pm}=\pm f_i \uvec{n}_i$} are exerted asymmetrically onto the fluid.
The sphere is transported by the resulting fluid flow (streamlines are shown, with dark (red) line segments corresponding to high magnitudes and light (yellow) ones to low magnitudes of the local fluid flow) for~$\alpha_i \neq 1/2$.  
A dashed circle of diameter~$\sigma_i$ indicates the effective swimmer size due to steric interactions between the swimmers.
(a) For~$f_i>0$, a pusher microswimmer is constructed, which expels fluid along its symmetry axis and draws fluid in from the sides.
(b) For a puller microswimmer ($f_i<0$), the directions of the fluid flow are inverted.
}
\label{fig:model}
\end{figure}

This said, we now detail our minimal microswimmer model, see Fig.~\ref{fig:model}, referring to one swimmer labeled by~$i$.
In this model, a no-slip boundary encloses the spherical swimmer body, the latter being centered at position~$\vect{r}_i$ and being of radius~$a_i$.
Below,~$\vect{v}_i$ and~$\vgr{\omega}_i$ denote the velocity and angular velocity of the sphere, respectively.

Additionally, two oppositely-oriented forces
\begin{equation}
 \vect{f}_{i\pm} = \pm f_i \, \uvec{n}_i
\label{swim_forces}
\end{equation}
of equal magnitude 
are exerted by the swimmer onto the surrounding fluid
at positions 
\begin{align}
 \vect{r}_{i+} = &\,\vect{r}_i + \alpha_i \, L_i \, \uvec{n}_i, \label{swim_force_rplus} \\ 
 \vect{r}_{i-} = &\,\vect{r}_i - (1-\alpha_i) \, L_i \, \uvec{n}_i,
\label{swim_force_rminus}
\end{align}
respectively, relative to its body center.
They move and rotate along with the sphere and
create the fluid flow that (self-)propels the swimmer.
Here,~$\uvec{n}_i$ is the unit vector describing the orientation of the axially symmetric swimmer,
$L_i > 2 a_i$ is the distance between the two force centers, 
and 
$|f_i|$ sets the magnitude of the forces.
Depending on the sign of~$f_i$, either pusher ($f_i>0$) or puller ($f_i<0$) microswimmers are constructed.
Furthermore, the real number~$\alpha_i$ (with~$a_i / L_i < \alpha_i < 1 / 2$) is a geometric parameter, see Fig.~\ref{fig:model}, that quantifies the breaking of the front--rear symmetry, which implies self-propulsion.
The swimmer self-propels in the direction of~$\uvec{n}_i$ for pushers, see Fig.~\ref{fig:model}(a), and~${}-\uvec{n}_i$ for pullers, see Fig.~\ref{fig:model}(b).

Moreover, an isotropic steric interaction between the swimmers is assumed that avoids unphysical overlap between force centers and bodies of different swimmers.
As indicated in Fig.~\ref{fig:model} and further detailed later, the effective center-to-center range of the steric interactions is denoted by~$\sigma_i$.

Next, we specify the hydrodynamic interactions in a system of~$N$ potentially different such model swimmers, labeled by~$i=1,\dots,N$.
For shorter notation, we furthermore define the phase-space coordinate~$\vect{X}_i = \left\{\vect{r}_i, \uvec{n}_i\right\}$ of each swimmer~$i$. 
In our overdamped system of microswimmers in suspension,
~$\vect{v}_i$ and~$\vgr{\omega}_i$ follow instantaneously from the microstate~$\vect{X}^N=\left\{\vect{X}_1,\dots,\vect{X}_N\right\}$.

In principle, hydrodynamic interactions are many-body interactions.\cite{Happel_Brenner,Dhont,Kim_Karrila}
Yet, already the lowest-order contributions beyond pairwise interactions are of fourth order in the ratio of body size to swimmer distance\cite{Dhont} and can be neglected when one is primarily interested in the effect of far-field hydrodynamic interactions, e.g., in semi-dilute suspensions.\cite{beenakker1983self,beenakker1984diffusion,nagele1996dynamics, nagele1997long, banchio2006many}
This is further supported by our use of repulsive steric interactions between swimmers, as detailed below, that keep them at distances from each other that are significantly larger than their hydrodynamic radii $a_i$, see also Fig.~\ref{fig:model}.
Thus, here we only account for pairwise interactions and restrict ourselves to an expansion up to (including) the third order, also known as the Rotne-Prager level.\cite{Rotne_1969_JCP,reichert2004hydrodynamic}

Following this idea,~$\vect{v}_i$ and~$\vgr{\omega}_i$ are connected to the (non-hydrodynamic) forces~$\vect{F}_j$ and torques~$\vect{T}_j$ acting on the swimmer bodies~$j=1,\dots,N$ and the self-propulsion forces that the swimmers exert via \cite{menzel2016dynamical}
\begin{align}
\left[
\hspace{-0.8 ex}
\begin{array}{c}
 \vect{v}_{i} \\[0.1cm]
 \vgr{\omega}_{i}
\end{array}
\hspace{-0.8 ex}
\right]
=
    \sum_{j=1}^{N}
    \Bigg( 
\left[
\hspace{-0.8 ex}
    \begin{array}{cc}
    \vgr{\mu}^{\mathrm{tt}}_{ij} & \vgr{\mu}^{\mathrm{tr}}_{ij}\\[0.1cm]
    \vgr{\mu}^{\mathrm{rt}}_{ij} & \vgr{\mu}^{\mathrm{rr}}_{ij}    
    \end{array}
    \hspace{-0.8 ex}
\right]
    &\cdot
\left[
\hspace{-0.8 ex}
    \begin{array}{c}
        \vect{F}_{j} \\[0.1cm]
        \vect{T}_{j}
    \end{array}
    \hspace{-0.8 ex}
\right] 
    +
\left[
\hspace{-0.8 ex}
    \begin{array}{cc}
    \vgr{\Lambda}^{\mathrm{tt}}_{ij} & \vect{0}\\[0.1cm]
    \vgr{\Lambda}^{\mathrm{rt}}_{ij} & \vect{0}
    \end{array}
    \hspace{-0.8 ex}
\right]
    \cdot
\left[
\hspace{-0.8 ex}
    \begin{array}{c}
        f \uvec{n}_{j} \\[0.1cm]
        \vect{0}
    \end{array}
    \hspace{-0.8 ex}
\right]
    \Bigg).
  \label{mobility_og}
\end{align}
Here, the mobility tensors representing the passive hydrodynamic interactions between two swimmer bodies~$i\neq j$ are given by \cite{Dhont,Rotne_1969_JCP,reichert2004hydrodynamic,menzel2016dynamical}
\begin{align}
\bm{\mu}_{ij}^{tt}&=\frac{1}{6\pi\eta} \bigg(\frac{3}{4r_{ij}}\Big({\bf {1}}+{{\mathbf{\hat r}_{ij}\mathbf{\hat r}_{ij}}}\Big)  \nonumber\\ &{}
+\frac{a^2_i+a^2_j}{4}\Big(\frac{1}{r_{ij}}\Big)^3\Big({\bf {1}}-3 \, {{\mathbf{\hat r}_{ij}\mathbf{\hat r}_{ij}}}\Big)\bigg), 
\label{mu_tt} \\ 
\bm{\mu}_{ij}^{rr}&={}-\frac{1}{8\pi\eta}\frac{1}{2}\left(\frac{1}{r_{ij}}\right)^3\left({\bf {1}}-3 \, {{\mathbf{\hat r}_{ij}\mathbf{\hat r}_{ij}}}\right), \label{mu_rr}\\ 
\bm{\mu}_{ij}^{tr}&=\bm{\mu}_{ij}^{rt}=\frac{1}{8\pi\eta}\left(\frac{1}{r_{ij}}\right)^3{ {\mathbf r_{ij}}}\times, 
\label{mu_tr}
\end{align} 
where~$\eta$ is the dynamic viscosity of the fluid, ``$\times$'' denotes the outer vector product, $\bm{1}$ represents the identity matrix, \mbox{$\vect{r}_{ij}=\vect{r}_j-\vect{r}_i$} is the distance vector, \mbox{$r_{ij}=|\vect{r}_{ij}|$} denotes its absolute value, and~$\uvec{r}_{ij}=\vect{r}_{ij} / r_{ij}$ is the corresponding unit vector.
Additionally, the passive ``self'' (i.e.,~$i=j$) mobilities read (no summation over repeated indices in these expressions)
\begin{equation}
 \bm{\mu}^\mathrm{tt}_{ii} = \mu^\mathrm{t}_i \, \bm{1}, \quad \bm{\mu}^\mathrm{rr}_{ii} = \mu^\mathrm{r}_i \, \bm{1}, \quad \bm{\mu}^\mathrm{tr}_{ii} =  \bm{\mu}^\mathrm{rt}_{ii} = \bm{0},
\end{equation}
with 
\begin{equation}
 \mu^\mathrm{t}_i=1/(6\pi\eta a_i), \quad \mu^\mathrm{r}_i=1/(8\pi\eta a^3_i).
\label{mu_self}
\end{equation}
Next, the active contribution to Eq.~(\ref{mobility_og}) is given by the tensors \cite{menzel2016dynamical}
\begin{eqnarray}
\bm{\Lambda}_{ij}^{\mathrm{tt}} &=& \bm{\mu}_{ij}^{\mathrm{tt}+}-\bm{\mu}_{ij}^{\mathrm{tt}-},
\label{Lambda_tt} \\
\bm{\Lambda}_{ij}^{\mathrm{rt}} &=& \bm{\mu}_{ij}^{\mathrm{rt}+}-\bm{\mu}_{ij}^{\mathrm{rt}-}, 
\label{Lambda_rt}
\end{eqnarray}
with
\begin{align}
\bm{\mu}_{ij}^{\mathrm{tt}\pm}&=
\frac{1}{8\pi\eta r_{ij}^{\pm}}\left({\bf {1}}+\mathbf{\hat r}_{ij}^{\pm}\mathbf{\hat r}_{ij}^{\pm}\right) 
{}+\frac{a_i^2}{24\pi\eta \left({r_{ij}^{\pm}}\right)^3}\left({\bf {1}}-3 \, \mathbf{\hat r}^{\pm}_{ij}\mathbf{\hat r}^{\pm}_{ij}\right), 
\label{mu_tt_pm}\\
\bm{\mu}_{ij}^{\mathrm{rt}\pm} &= \frac{1}{8\pi\eta \left({r_{ij}^{\pm}}\right)^3} \, \mathbf r_{ij}^{\pm}\times,
\label{mu_rt_pm}
\end{align} 
and 
\begin{eqnarray}
\mathbf r_{ij}^+ &=& \mathbf r_{ij}+\alpha_j \, L_j \, \mathbf{\hat n}_j, \label{defplus}
\\
\mathbf r_{ij}^- &=& \mathbf r_{ij}-(1-\alpha_j) \, L_j \, \mathbf{\hat n}_j. \label{defminus}
\end{eqnarray}
As can be seen, there is only little change to the one-species case ($a_i = a_j \equiv a$) \cite{menzel2016dynamical} at this order of the expansion in~$a_{k} / r_{ij}$,~$k=i,j$, namely only in Eq.~(\ref{mu_tt}).

Setting~$i=j$ in Eqs.~(\ref{Lambda_tt}) and (\ref{Lambda_rt}), the velocity and angular velocity of a free swimmer~$i$ are obtained as \cite{pessot2018} 
\begin{align}
 \!\!\vect{v}_{0i} &= \frac{a_i}{2 L_i} \frac{1-2\alpha_i}{\alpha_i (1-\alpha_i)} \left( 3 - \frac{a_i^2}{L_i^2} \frac{1-\alpha_i+\alpha_i^2}{\alpha_i^2(1-\alpha_i)^2} \right) \mu^\mathrm{t}_i f_i \, \uvec{n}_i
\label{v0}
\end{align}
and, respectively,~$\vgr{\omega}_{0i} = \vect{0}$.
Thus, in the absence of thermal noise and outer influences, this kind of swimmer self-propels along a straight trajectory.
Corresponding circle swimmers of axial asymmetry and a non-vanishing~$\vgr{\omega}_{0i}$ were considered in a previous work.\cite{hoell2017dynamical}

We remark that, for simplicity, we here do not account for the distortions caused by the finite spherical swimmer bodies on the flow field induced by the active force centers.\cite{Kim_Karrila, adhyapak2017flow}
That is, when discussing the active mobility tensors $\vgr{\Lambda}_{ij}^{\mathrm{tt}}$ and $\vgr{\Lambda}_{ij}^{\mathrm{rt}}$
 for $i\neq j$, in effect we only consider terms in~$a_j / L_j$ to leading order. 

Finally, the forces and torques in Eq.~(\ref{mobility_og}) remain to be defined.
First, we set the overall potential in our system of~$N$ swimmers as
\begin{equation}
 U(\vect{r}_1,\dots,\vect{r}_N) = \sum\limits_{k=1}^{N} u^k_\mathrm{ext} (\vect{r}_k) +\!\! \!\! \!\!\sum\limits_{k, \, l=1; \, k < l}^N \!\! \!\! \!\! u^{kl}(\revi{|\vect{r}_k-\vect{r}_l|}).
\label{potential}
\end{equation}
Here, the external potentials~$u^k_\mathrm{ext}$ can differ for different particles~$k$.
Additionally, pairwise steric interactions are introduced via~$u^{kl}(\revi{|\vect{r}_k-\vect{r}_l|})$, 
which we specify for the applications in Sec.~\ref{sec:results} as the GEM-4 potential \cite{mladek2006formation,archer2014solidification}
\begin{equation}
 u^{kl}(\revi{|\vect{r}_k-\vect{r}_l|}) = \epsilon^{kl}_0 \exp\left( - \left(\frac{|\vect{r}_k-\vect{r}_l|}{\sigma_{kl}}\right)^4 \right),
\label{interaction_potential}
\end{equation}
with the potential strength~$\epsilon^{kl}_0$ and the effective diameter~$\sigma_{kl}=(\sigma_k+\sigma_l)/2$ being set for each pair~$k$ and~$l$.

The forces~$\vect{F}_j$ in Eq.~(\ref{mobility_og}) then read 
\begin{equation}
 \vect{F}_j = {}- k_\mathrm{B} T \, \nabla_{\vect{r}_j} \ln P - \nabla_{\vect{r}_j} U(\vect{r}_1, \dots, \vect{r}_N),
\end{equation}
where the effect of thermal forces enters via the first term based on the effective entropic potential,\cite{doi1988theory}
which involves the microstate probability density~$P=P(\vect{X}^N,t)$, the Boltzmann factor~$k_\mathrm{B}$, and the temperature~$T$ of the system.
This expression ensures that the correct (translational) diffusion terms eventually appear in the statistical description in Sec.~\ref{sec:ddft}.

Similarly, the torques in Eq.~(\ref{mobility_og}) are given by
\begin{equation}
 \vect{T}_j = {}- k_\mathrm{B} T \, \uvec{n}_j \times \nabla_{\uvec{n}_j} \ln P.
\end{equation}
Again, this expression correctly reproduces (rotational) diffusion in the statistical description, see Sec.~\ref{sec:ddft}.

\section{Derivation of the dynamical density functional theory}
\label{sec:ddft}

In this section, we derive the partial differential equations describing the dynamical microscopic statistics of a multi-component microswimmer system via dynamical density functional theory (DDFT), building on the derivations of the one-component case.\cite{menzel2016dynamical, hoell2017dynamical} 
For this purpose, the hydrodynamic swimmer model overviewed in Sec.~\ref{sec:model} is used as an input.
The resulting theory covers, combines, and extends several previously considered theories for systems of,
e.g., one-species microswimmer suspensions,\cite{menzel2016dynamical}
``dry'' --- i.e., not hydrodynamically-interacting --- self-propelled particles,\cite{wensink2008aggregation,wittkowski2011dynamical}
 hydrodynamically interacting passive colloidal particles,\cite{rex2009dynamical}
and binary mixtures of dry passive colloidal particles.\cite{archer2005dynamical}

First, we specify our system, 
which contains two different species of microswimmers suspended in a surrounding bulk fluid.
For these species, the number of corresponding swimmers in the system is given by~$N_\mathrm{A}$ and~$N_\mathrm{B}$, respectively,
adding up to a total of~$N=N_\mathrm{A}+N_\mathrm{B}$ swimmers.
Here, we order the swimmers by species, such that swimmers~$1,\dots,N_\mathrm{A}$ belong to the first species and swimmers~$N_\mathrm{A}+1,\dots,N$ to the second species.
Additionally, a constant temperature~$T$ of the fluid and a constant volume of the system are assumed.
We adhere to the swimmer model introduced in Sec.~\ref{sec:model},
with all swimmers of species~$\nu\in\{\mathrm{A},\mathrm{B}\}$ featuring the same parameters~$a_\nu$, $f_\nu$, $\alpha_\nu$, $L_\nu$, and~$\sigma_\nu$.
Setting~$f_\nu = 0$, also passive particles can be described accordingly, i.e., active--passive mixtures are covered by our theory as well.

Our starting point to derive the statistical description is the 
many-body Smoluchowski equation \cite{doi1988theory}
\begin{equation}
\frac{\partial P}{\partial t}= {}-  \sum_{i=1}^{N}\Big(\nabla_{\!\vect{r}_{i}} \cdot\left(\mathbf v_{i} P \right)
  +\left(\uvec{n}_{i} \times \nabla_{\!\uvec{n}_{i} }\right)\cdot\left(\bm{\omega}_{i} P\right)\Big)
\label{Smoluchowski}
\end{equation}
for the overdamped dynamics of our microswimmers.
Here,~$P=P(\vect{X}_1,...,\vect{X}_N,t)$ 
denotes the microstate probability density of the corresponding configuration at time~$t$.
The velocities~$\vect{v}_i$ and the angular velocities~$\vgr{\omega}_i$ are again related to the forces, torques, and the self-propulsion mechanisms via Eq.~(\ref{mobility_og}).

Next, we introduce~$\vect{X}^{m}_\mathrm{A}=\{ \vect{X}_{1}, \dots , \vect{X}_{m} \}$ and \mbox{$\vect{X}^{n}_\mathrm{B}=\{ \vect{X}_{N_\mathrm{A}+1}, \dots , \vect{X}_{N_\mathrm{A}+n} \}$} as short notations for the sets containing the phase-space coordinates of the first~$m$ swimmers of species~A and, respectively, the first~$n$ swimmers of species~B in the system.
Since all swimmers are identical, we now define, for~$m \leq N_\mathrm{A}$ and~$n \leq N_\mathrm{B}$,
the reduced~$(m, n)$-swimmer density 
$\rho^{(m,n)}(\vect{X}^m_\mathrm{A}, \vect{X}^n_\mathrm{B},t)$
of finding (any)~$m$~swimmers of species~A
and (any) $n$~swimmers of species~B
at the coordinates indicated in the argument.
It is obtained from the full probability distribution~$P(\vect{X}^{N_\mathrm{A}}_\mathrm{A}, \vect{X}^{N_\mathrm{B}}_\mathrm{B},t)$ by integrating out the degrees of freedom of \mbox{$N_\mathrm{A}-m$} swimmers of species~A and \mbox{$N_\mathrm{B}-n$} swimmers of species~B,
reading
\begin{widetext} 
\vspace{-10 pt}
\begin{align}
\rho^{(m,n)}(\vect{X}^m_\mathrm{A}, \vect{X}^n_\mathrm{B},t) = &\frac{N_\mathrm{A}!}{(N_\mathrm{A}-m)!} \, \frac{N_\mathrm{B}!}{(N_\mathrm{B}-n)!}
\int \! \! \mathrm{d}\vect{X}_{m+1} \, \, \dots \int \! \! \mathrm{d}\vect{X}_{N_\mathrm{A}}  \int \! \! \mathrm{d}\vect{X}_{N_\mathrm{A}+n+1} \, \, \dots \int \! \! \mathrm{d}\vect{X}_{N_\mathrm{A}+N_\mathrm{B}} \, P(\vect{X}^{N_\mathrm{A}}_\mathrm{A}, \vect{X}^{N_\mathrm{B}}_\mathrm{B},t).
\label{nbody}
\end{align}
\vspace{-10 pt}
\end{widetext}
Here, the prefactors result from the considered indistinguishability between swimmers of the same species.
Particularly, we define the one-swimmer densities \mbox{$\rho_\mathrm{A}(\vect{X},t):=\rho^{(1,0)}(\vect{X},t)$} and
\mbox{$\rho_\mathrm{B}(\vect{X},t):=\rho^{(0,1)}(\vect{X},t)$}.
Instead of referring to one specific swimmer, the coordinates~$\vect{X}$ now simply identify ``a swimmer'' of the corresponding species.
Furthermore, reduced densities with~$m+n=2$ ($m+n=3$) will be referred to as \mbox{two-swimmer} (three-swimmer) densities below.  

Our aim is to derive a physically well-grounded, closed set of coupled dynamical equations for the two one-swimmer densities.
Thus, eventually, there shall be no remaining explicit dependence on the (generally unknown) higher-order densities.
The starting point for our derivation is the many-body Smoluchowski equation given in Eq.~(\ref{Smoluchowski}). 
We first integrate out all swimmer coordinates except for those of one swimmer of species~A.
Second, we integrate out in the initial Eq.~(\ref{Smoluchowski}) all swimmer coordinates except for those of one swimmer of species~B.
As a result, we obtain one dynamical equation for~$\rho_\mathrm{A}(\vect{X},t)$ and one for~$\rho_\mathrm{B}(\vect{X},t)$, respectively. 
These equations (given below) form a coupled set, but at this point still contain higher-order densities and thus require an additional closure,
as will be addressed afterwards via methods of dynamical density functional theory.

\allowdisplaybreaks
The corresponding equation for species~A reads
\begin{align}
\frac{\partial\rho_\mathrm{A}(\vect{X},t)}{\partial t} = 
&-\nabla_{\vect{r}}\cdot\Big(\curr{\mathrm{A}}^\mathrm{tt}\!+\!\curr{\mathrm{A}}^\mathrm{tr}\!
+\!\curr{\mathrm{A}}^\mathrm{ta}\! 
	+ \!\currk{\mathrm{AA}}^\mathrm{tt} \! \notag \\&+ \! \currk{\mathrm{AA}}^\mathrm{tr} \!
+ \!\currk{\mathrm{AA}}^\mathrm{ta} \!
	+ \!\currk{\mathrm{AB}}^\mathrm{tt} \!+ \!\currk{\mathrm{AB}}^\mathrm{tr} \!+ 
\! \currk{\mathrm{AB}}^\mathrm{ta}\Big) \notag \\
&-(\mathbf{\hat n} \times \nabla_{\mathbf{\hat n}})\cdot\Big(\curr{\mathrm{A}}^\mathrm{rt}\!+\!
\curr{\mathrm{A}}^\mathrm{rr}\!
+\!\curr{\mathrm{A}}^\mathrm{ra}\! 
	+ \!\currk{\mathrm{AA}}^\mathrm{rt} \notag \\ & +\!\currk{\mathrm{AA}}^\mathrm{rr} \!
+\!\currk{\mathrm{AA}}^\mathrm{ra}\!
	+ \!\currk{\mathrm{AB}}^\mathrm{rt} \!+\! \currk{\mathrm{AB}}^\mathrm{rr} \!
+ \!\currk{\mathrm{AB}}^\mathrm{ra}\Big).
\label{BBGKY1}
\end{align}
In this expression, the current densities labeled as $\curr{\cdot}^{\cdot \cdot}$ do not involve hydrodynamic interactions between swimmers.
These current densities are given by
\begin{align}
\curr{\mathrm{A}}^\mathrm{tt}
=&
{}-\mu^{\mathrm{t},\mathrm{A}} \Big(k_\mathrm{B} T \, \nabla_{\mathbf r}\,\rho_\mathrm{A}(\vect{X},t)+\rho_\mathrm{A}(\vect{X},t) \nabla_{\mathbf r}\,u^\mathrm{A}_\mathrm{ext}(\mathbf r) \nonumber \\ &
+\! \int \! \! \mathrm{d}\vect{X}' \rh{2,0}(\vect{X}, \vect{X}', t)\nabla_{\vect{r}}u^\mathrm{AA}(\revi{|\vect{r}-\vect{r}'|}) \nonumber\\ &
 +\! \int \! \! \mathrm{d}\vect{X}' \rh{1,1}(\vect{X}, \vect{X}', t)\nabla_{\vect{r}}u^\mathrm{AB}(\revi{|\vect{r}-\vect{r}'|}) \Big),
\label{eqJ1}  \\
\curr{\mathrm{A}}^\mathrm{ta}
=&
f_\mathrm{A}{\bm{\Lambda}^{\mathrm{tt},\mathrm{AA}}_{\vect{r},\vect{X}}}\cdot\mathbf{\hat n} \, \rho_\mathrm{A}(\vect{X}, t)   
,
\label{eqJ3} \\
\curr{\mathrm{A}}^\mathrm{rr}
=&
{}-k_\mathrm{B} T \, \mu^{\mathrm{r},\mathrm{A}} \, \uvec{n}\times \nabla_{\uvec{n}} \, \rho_\mathrm{A}(\vect{X}, t), \label{eqJ5} 
\\ \curr{\mathrm{A}}^\mathrm{tr}=& \, \curr{\mathrm{A}}^\mathrm{rt}=\curr{\mathrm{A}}^\mathrm{ra}=\vect{0}. \label{eqJ_rest}
\end{align}
In contrast to that, the current densities involving hydrodynamic interactions between pairs of swimmers of species~A follow as
\begin{align}
\!\!\currk{\mathrm{AA}}^\mathrm{tt}\!
=
&-\! \int \! \! \mathrm{d}\vect{X}' \,{\vgr{\mu}^{\mathrm{tt},\mathrm{AA}}_{\vec{r}, \vec{r}'}}\cdot
\bigg(k_\mathrm{B} T \, \nabla_{\vect{r}'}\rh{2,0}( \vect{X},\vect{X}',t) 
\notag\\ &
+\rh{2,0}( \vect{X},\vect{X}',t)\nabla_{\vect{r}'}\left(u^\mathrm{A}_\mathrm{ext}(\vect{r}')+u^\mathrm{AA}(\revi{|\vect{r}-\vect{r}'|})\right)\notag\\ &
+ \! \int \! \! \mathrm{d}\vect{X}'' \rh{2,1}(\vect{X},\vect{X}',\vect{X}'', t)\nabla_{\vect{r}'} u^\mathrm{AB}(\revi{|\vect{r}'-\vect{r}''|}) \notag\\ &
+ \! \int \! \! \mathrm{d}\vect{X}'' \rh{3,0}(\vect{X},\vect{X}',\vect{X}'', t)\nabla_{\vect{r}'} u^\mathrm{AA}(\revi{|\vect{r}'-\vect{r}''|}) \bigg),
\label{eqKA1} \\ 
\!\!\currk{\mathrm{AA}}^\mathrm{tr}\!
=&
{}-\! \int \! \! \mathrm{d}\vect{X}' \, k_\mathrm{B} T \, \vgr{\mu}_{\vect{r}, \vect{r}'}^{\mathrm{tr},\mathrm{AA}} 
\, (\uvec{n}'\times\nabla_{\uvec{n}'})\rh{2,0}( \vect{X},\vect{X}',t)\nonumber\\=&
\, \vec{0},
\label{eqKA2} \\ 
\!\!\currk{\mathrm{AA}}^\mathrm{ta}\!
=&
f_\mathrm{A} \! \int \! \! \mathrm{d}\vect{X}' \,{\vgr{\Lambda}^{\mathrm{tt},\mathrm{AA}}_{\vect{r},\vect{X}'}}\cdot\uvec{n}'\rh{2,0}( \vect{X},\vect{X}',t),
\label{eqKA3} \\ 
\!\!\currk{\mathrm{AA}}^\mathrm{rt}\!
=
&-\! \int \! \! \mathrm{d}\vect{X}' \,{\vgr{\mu}^{\mathrm{rt},\mathrm{AA}}_{\vect{r}, \vect{r}'}}
 \bigg(k_\mathrm{B} T \, \nabla_{\vect{r}'}\rh{2,0}( \vect{X},\vect{X}',t)
\notag\\ &
+\rh{2,0}(\vect{X},\vect{X}',t) \nabla_{\vect{r}'} \left(u^\mathrm{A}_\mathrm{ext}(\vect{r}')
+ u^\mathrm{AA}(\revi{|\vect{r}-\vect{r}'|}) \right)\notag\\ &
+ \! \int \! \! \mathrm{d}\vect{X}'' \rh{2,1}(\vect{X},\vect{X}',\vect{X}'', t)\nabla_{\vect{r}'} u^\mathrm{AB}(\revi{|\vect{r}'-\vect{r}''|}) \notag\\ &
+ \! \int \! \! \mathrm{d}\vect{X}'' \rh{3,0}(\vect{X},\vect{X}',\vect{X}'', t)\nabla_{\vect{r}'} u^\mathrm{AA}(\revi{|\vect{r}'-\vect{r}''|}) \bigg),
\label{eqKA4}\\ 
\!\!\currk{\mathrm{AA}}^\mathrm{rr}\!
=&
- \! \int \! \! \mathrm{d}\vect{X}' \, k_\mathrm{B} T \, \vgr{\mu}^{\mathrm{rr},\mathrm{AA}}_{\vect{r}, \vect{r}'}
\cdot (\uvec{n}'\times\nabla_{\uvec{n}'})\rh{2,0}(\vect{X},\vect{X}',t)\nonumber\\=& \,
\vec{0},
\label{eqKA5} \\ 
\!\!\currk{\mathrm{AA}}^\mathrm{ra}\!
=&
f_\mathrm{A} \! \int \! \! \mathrm{d}\vect{X}' \,{\vgr{\Lambda}^{\mathrm{rt},\mathrm{AA}}_{\vect{r}, \vect{X}'}}
\, \uvec{n}' \rh{2,0}(\vect{X},\vect{X}',t).
\label{eqKA6}
\end{align}
Third, the current densities associated with hydrodynamic effects of swimmers of species~B on swimmers of species~A are
\allowdisplaybreaks[0]
\begin{align}
\!\!\currk{\mathrm{AB}}^\mathrm{tt}\!
=
&-\! \int \! \! \mathrm{d}\vect{X}' \,{\vgr{\mu}^{\mathrm{tt},\mathrm{AB}}_{\vec{r}, \vec{r}'}}\cdot
\bigg(k_\mathrm{B} T \, \nabla_{\vect{r}'}\rh{1,1}( \vect{X},\vect{X}',t) \nonumber\\ &
+\rh{1,1}( \vect{X},\vect{X}',t)\nabla_{\vect{r}'} \left(u^\mathrm{B}_\mathrm{ext}(\vect{r}')
+u^\mathrm{AB}(\revi{|\vect{r}-\vect{r}'|})\right) \nonumber\\ &
+ \! \int \! \! \mathrm{d}\vect{X}'' \rh{1,2}(\vect{X},\vect{X}',\vect{X}'', t)\nabla_{\vect{r}'} u^\mathrm{BB}(\revi{|\vect{r}'-\vect{r}''|}) \nonumber\\ &
+ \! \int \! \! \mathrm{d}\vect{X}'' \rh{2,1}(\vect{X},\vect{X}'',\vect{X}', t)\nabla_{\vect{r}'} u^\mathrm{AB}(\revi{|\vect{r}'-\vect{r}''|}) \bigg),
\label{eqK1} \\ 
\!\!\currk{\mathrm{AB}}^\mathrm{tr}\!
=&
{}-\! \int \! \! \mathrm{d}\vect{X}' \, k_\mathrm{B} T \, \vgr{\mu}_{\vect{r}, \vect{r}'}^{\mathrm{tr},\mathrm{AB}} 
\, (\uvec{n}'\times\nabla_{\uvec{n}'})\rh{1,1}( \vect{X},\vect{X}',t)\nonumber\\=& \,
\vec{0},
\label{eqK2} \\ 
\!\!\currk{\mathrm{AB}}^\mathrm{ta}\!
=&
f_\mathrm{B} \! \int \! \! \mathrm{d}\vect{X}' \,{\vgr{\Lambda}^{\mathrm{tt},\mathrm{AB}}_{\vect{r},\vect{X}'}}\cdot\uvec{n}'\rh{1,1}( \vect{X},\vect{X}',t),
\label{eqK3} \\ 
\!\!\currk{\mathrm{AB}}^\mathrm{rt}\!
=
&-\! \int \! \! \mathrm{d}\vect{X}' \,{\vgr{\mu}^{\mathrm{rt},\mathrm{AB}}_{\vect{r}, \vect{r}'}}
 \bigg(k_\mathrm{B} T \, \nabla_{\vect{r}'}\rh{1,1}( \vect{X},\vect{X}',t)
\nonumber\\ &
+\rh{1,1}(\vect{X},\vect{X}',t) \nabla_{\vect{r}'} \left( u^\mathrm{B}_\mathrm{ext}(\vect{r}')
+ u^\mathrm{AB}(\revi{|\vect{r}-\vect{r}'|}) \right) \nonumber\\ &
+ \! \int \! \! \mathrm{d}\vect{X}'' \rh{1,2}(\vect{X},\vect{X}',\vect{X}'', t)\nabla_{\vect{r}'} u^\mathrm{BB}(\revi{|\vect{r}'-\vect{r}''|}) \nonumber\\ &
+ \! \int \! \! \mathrm{d}\vect{X}'' \rh{2,1}(\vect{X},\vect{X}'',\vect{X}', t)\nabla_{\vect{r}'} u^\mathrm{AB}(\revi{|\vect{r}'-\vect{r}''|}) \bigg),
\label{eqK4} \\ 
\!\!\currk{\mathrm{AB}}^\mathrm{rr}\!
=&
- \! \int \! \! \mathrm{d}\vect{X}' \, k_\mathrm{B} T \, \vgr{\mu}^{\mathrm{rr},\mathrm{AB}}_{\vect{r}, \vect{r}'}
\cdot (\uvec{n}'\times\nabla_{\uvec{n}'})\rh{1,1}(\vect{X},\vect{X}',t)\nonumber\\=& \,
\vec{0},
\label{eqK5} \\ 
\!\!\currk{\mathrm{AB}}^\mathrm{ra}\!
=&
f_\mathrm{B} \! \int \! \! \mathrm{d}\vect{X}' \,{\vgr{\Lambda}^{\mathrm{rt},\mathrm{AB}}_{\vect{r}, \vect{X}'}}
\, \uvec{n}' \rh{1,1}(\vect{X},\vect{X}',t).
\label{eqK6}
\end{align}
Here, the tensors $\vgr{\mu}^{\cdot \cdot}_{\cdot \cdot}$ and $\vgr{\Lambda}^{\cdot \cdot}_{\cdot \cdot}$ follow from the definitions in Eqs.~(\ref{mu_tt})--(\ref{defminus}) by inserting the parameters corresponding to the (phase-space) coordinates given in the subscripts and the combination of species referred to in the superscripts.
The current densities $\currk{\mathrm{AA}}^\mathrm{tr}$, $\currk{\mathrm{AA}}^\mathrm{rr}$, $\currk{\mathrm{AB}}^\mathrm{tr}$, and $\currk{\mathrm{AB}}^\mathrm{rr}$ vanish for spherical swimmer bodies
because the corresponding mobility tensors are independent of $\uvec{n}'$, see Eqs.~(\ref{mu_rr}) and~(\ref{mu_tr}).
Integrating the remaining gradient expressions over the closed surface of the unit sphere yields zero in each case.
For non-spherical swimmer bodies, however, these current densities (as well as $\curr{\mathrm{A}}^\mathrm{tt}$, $\curr{\mathrm{A}}^\mathrm{rt}$, and~$\curr{\mathrm{A}}^\mathrm{ra}$) could be non-zero.
Moreover, we remark that all $\currk{}$'s become zero if hydrodynamic interactions are neglected.

An analogous dynamical equation for $\rho_\mathrm{B}(\vect{X},t)$ follows by replacing 
$\mathrm{A}\to\mathrm{B}$, $\mathrm{B}\to\mathrm{A}$, and $\rh{m,n}\to\rh{n,m}$.
Moreover, because of our convention of ordering species coordinates by first A and then B, we need to replace 
\begin{align*}
 \rh{1,1}(\vect{X},\vect{X}',t)&\to\rh{1,1}(\vect{X}',\vect{X},t),\\ 
\rh{1,2}(\vect{X},\vect{X}',\vect{X}'',t)&\to\rh{2,1}(\vect{X}',\vect{X}'',\vect{X},t),\\ 
\rh{2,1}(\vect{X},\vect{X}',\vect{X}'',t)&\to\rh{1,2}(\vect{X}'',\vect{X},\vect{X}',t),\\ 
\rh{2,1}(\vect{X},\vect{X}'',\vect{X}',t)&\to\rh{1,2}(\vect{X}',\vect{X},\vect{X}'',t).
\end{align*}

Obviously, the above equations (\ref{eqJ1})--(\ref{eqK6}) depend on unknown higher-order densities.
In principle, one can now find dynamical equations for these quantities by applying corresponding integral operations on Eq.~(\ref{Smoluchowski}),
but the resulting equations again contain unknown densities of even higher order.
This escalating loop is typical for BBGKY-like hierarchies\cite{hansen1990theory} and must be truncated and closed 
by appropriate approximations of the higher-order densities,
e.g., as functions of the one-swimmer densities.
In the following, DDFT methods will be employed for this purpose.

The main step in DDFT \cite{evans1979nature,evans1992density,marconi1999dynamic,marconi2000dynamic,archer2004dynamical,chan2005time,espanol2009derivation,evans2010density,lowen2010density,wittkowski2011dynamical} is the \emph{adiabatic approximation}.
It transfers equilibrium closure relations established in (classical)  density functional theory (DFT) to the non-equilibrium case.
Particularly, DDFTs imply that the higher-order densities relax faster than the one-swimmer densities,\cite{espanol2009derivation}
as is conceivable for typical overdamped systems of colloidal particles (i.e., at low Reynolds numbers) and thus also for microswimmers.\cite{menzel2016dynamical}

In equilibrium, DFT states that each observed density profile results from exactly one, uniquely specified external potential working on the corresponding particles.\cite{singh1991density,evans1992density,chan2005time,lowen2010density,marconi1999dynamic,marconi2000dynamic,archer2004dynamical}
We call these potentials $\Phi^\nu_\mathrm{ext}(\vect{X})$, $\nu=\mathrm{A,B}$, for the two species in our case.
DDFT assumes these relations to hold at any time $t$.
Thus, the external DFT potentials become time-dependent, and we denote them by $\Phi^\nu_\mathrm{ext}(\vect{X},t)$.
We remark that the equilibrium relations strictly hold only for $f_\nu=0$, $\nu=\mathrm{A,B}$, i.e., for passive particles.
This limits the applicability of the theory when activity-induced correlation effects in the higher-order densities dominate the behavior of the system.
Nevertheless, the overdamped nature of the systems favors the DDFT approach.
Previously, bulk swimmer--swimmer pair distribution functions have been determined \cite{hoell2018particle} by combining DDFT with a Percus-like \cite{percus1962approximation} test-particle protocol.

We now discuss the above-introduced virtual external potentials, which may (and generally will) differ for the two species.
In contrast to the ``real'' external potential introduced in Eq.~(\ref{potential}), a dependence on the orientations of the swimmers here is allowed,
and indeed even needed when the distributions of the orientations become non-uniform.

It must be stressed that these virtual potentials do not need to be determined explicitly. 
Repeating usual steps in derivations of DDFTs, we will in the following show two different ways of expressing $\Phi^\nu_\mathrm{ext}(\vect{X},t)$ so that they can be eliminated from the mathematical description.
Accordingly, we obtain expressions that help us to close the above BBGKY-like set of equations.

We start from the equilibrium grand potential as a functional of the one-swimmer densities, which is minimal for the equilibrium density distributions.
The general ansatz for this functional can be written as\cite{archer2005dynamical}
\begin{equation}
\label{Omega}
\Omega\left[\rho_\mathrm{A},\rho_\mathrm{B}\right]
=
\! \! \! \! \sum\limits_{\nu=\mathrm{A},\mathrm{B}} \! \! \! \! \Big( \mathcal{F}^\nu_{\!\!\mathrm{ext}}\left[\rho_\nu\right]
+ \mathcal{F}^\nu_{\!\mathrm{id}}\left[\rho_\nu\right] \Big)
+ \mathcal{F}_{\!\!\mathrm{exc}}\left[\rho_\mathrm{A},\rho_\mathrm{B}\right].
\end{equation}
Here, all terms on the right-hand side except for the last one are known analytically. 
Namely,
\begin{equation}
\mathcal{F}^\nu_{\!\mathrm{id}}\left[\rho_\nu\right]
= k_\mathrm{B} T
\! \! \int \! \! \mathrm{d}\vect{X}\,\rho_\nu(\vect{X})
\Big(\ln\left(\lambda_\nu^3\rho_\nu(\vect{X})\right)-1\Big), 
\end{equation}
$\nu=\mathrm{A,B}$, is the ideal gas part, with $\lambda_\nu$ the corresponding thermal de Broglie wavelength $\lambda_\nu$.
The contributions due to the external DFT potentials read
\begin{equation}
{\mathcal{F}}^\nu_{\!\!\mathrm{ext}}\left[\rho_\nu \right] = \! \! \int \! \! \mathrm{d}\vect{X}\,\rho_\nu(\vect{X}) \, \Phi^\nu_{\mathrm{ext}}(\vect{X}),
\label{F_ext}
\end{equation}
$\nu=\mathrm{A,B}$.
For our purposes, we may assume the chemical potentials to be combined with the external potentials.

Finally, the third contribution $\mathcal{F}_{\!\!\mathrm{exc}}$ includes interactions between the particles 
and represents the excess free energy beyond the ideal gas part.
In almost all situations, an exact expression for $\mathcal{F}_{\!\!\mathrm{exc}}$ is not known analytically,
and it must be approximated by an appropriate functional depending on the case at hand.
Typically, this assumption needs to be carefully tested against experimental and simulational results. 
Nevertheless, the general theoretical framework up to this point applies to any interaction potential, here independent of the orientations of the swimmers
(in principle, this restriction could be lifted, e.g., when describing rod-like active particles\cite{Rex_2007_PRE, wensink2008aggregation}). 

In equilibrium, the actual magnitude of the grand potential is found by minimizing the grand potential functional over all possible density distributions.
Thus, the equilibrium density fields $\rho_{\nu}^{\mathrm{eq}}(\vect{X})$ satisfy
\begin{equation}
 0 = \left.\frac{\delta\Omega}{\delta \rho_\nu(\vect{X})}\right|_{\rho_\nu\equiv\rho_{\nu}^{\mathrm{eq}}}
\end{equation}
for $\nu=\mathrm{A},\mathrm{B}$.
Inserting Eqs.~(\ref{Omega})--(\ref{F_ext}) leads to
\begin{equation}
{}-\Phi^\nu_{\!\!\mathrm{ext}}(\vect{X})\,=  k_\mathrm{B} T \, \ln\left(\lambda^3_\nu \rho_{\nu}^{\mathrm{eq}}(\vect{X})\right)
+\left.\frac{\delta {\mathcal{F}}_{\!\!\mathrm{exc}}}{\delta \rho_\nu(\vect{X})}\right|_{\rho_\nu\equiv\rho_{\nu}^{\mathrm{eq}}} \qquad
\label{rhoeq_equilibrium_prev}
\end{equation}
for $\nu=\mathrm{A,B}$.

Second, we employ standard equilibrium statistical mechanics.\cite{Gray_Gubbins}
In equilibrium, the static system properties are set completely by the temperature and the overall potential 
\mbox{$U=U(\vect{X}_1,\dots,\vect{X}_N)$}
 as defined in Eq.~(\ref{potential}), writing $\Phi^\nu_\mathrm{ext}(\vect{X})$ instead of $u^\nu_\mathrm{ext}(\vect{r})$.
Thus, the microstate probability density is given by
\begin{equation}
 P \equiv P^\mathrm{eq} \propto \exp({}- \beta U),
\label{Boltzmann}
\end{equation}
where $\beta=(k_\mathrm{B} T)^{-1}$.

Applying the gradient with respect to the position of the first swimmer, which is of species~A,
leads to
\begin{align}
 \nabla_{\vect{r}_1} P^\mathrm{eq} = &{}-\beta P^\mathrm{eq} \Bigg( \nabla_{\vect{r}_1} \Phi^\mathrm{A}_\mathrm{ext}(\vect{X}_{1}) 
+ \nabla_{\vect{r}_1} \sum\limits_{j=2}^{N_\mathrm{A}} u^\mathrm{AA}(\revi{|\vect{r}_1-\vect{r}_j|}) \nonumber\\&+ \nabla_{\vect{r}_1} \! \! \! \! \!  \sum\limits_{j=N_\mathrm{A}+1}^{N_\mathrm{A}+N_\mathrm{B}} \! \! \! \! u^\mathrm{AB}(\revi{|\vect{r}_1-\vect{r}_j|}) \Bigg).
\end{align}
Since swimmers of the same species are considered to be identical and indistinguishable, we may write 
\begin{align}
 k_\mathrm{B} T \, \nabla_{\vect{r}} \rho_\mathrm{A}^\mathrm{eq}(\vect{X}) = &{}- \rho_\mathrm{A}^\mathrm{eq}(\vect{X}) \nabla_{\vect{r}} \Phi^\mathrm{A}_\mathrm{ext}(\vect{X}) 
  \nonumber\\&{}
  - \! \! \int \! \! \mathrm{d} \vect{X}' \, \rho^{(2,0),\mathrm{eq}}(\vect{X},\vect{X}') \nabla_{\vect{r}} u^\mathrm{AA}(\revi{|\vect{r}-\vect{r}'|}) \nonumber\\&{}
  - \! \! \int  \! \!\mathrm{d} \vect{X}' \, \rho^{(1,1),\mathrm{eq}}(\vect{X},\vect{X}') \nabla_{\vect{r}} u^\mathrm{AB}(\revi{|\vect{r}-\vect{r}'|})
\label{YBGA1}
\end{align}
after integrating over the coordinates of all but the first swimmer of species~A and using Eq.~(\ref{nbody}). 
This constitutes a lowest-order member of the binary-mixture translational \emph{Yvon-Born-Green (YBG) relations}.\cite{hansen1990theory,Gray_Gubbins}
Combining Eqs.~(\ref{rhoeq_equilibrium_prev}) and (\ref{YBGA1}), $\Phi^\mathrm{A}_\mathrm{ext}(\vect{X})$ is eliminated and 
\begin{align}
 &\! \!\int \! \!\mathrm{d} \vect{X}' \, \rh{2,0}(\vect{X},\vect{X}',t) \, \nabla_{\vect{r}} u^\mathrm{AA}(\revi{|\vect{r}-\vect{r}'|}) \nonumber \\
 + &\! \! \int \! \!\mathrm{d} \vect{X}' \, \rh{1,1}(\vect{X},\vect{X}',t) \, \nabla_{\vect{r}} u^\mathrm{AB}(\revi{|\vect{r}-\vect{r}'|}) \nonumber \\&= \rho_\mathrm{A}(\vect{X},t) \, \nabla_{\vect{r}} \frac{\delta {\mathcal{F}}_{\!\!\mathrm{exc}}}{\delta \rho_\mathrm{A}(\vect{X},t)} 
\label{ybg_a}
\end{align}
is obtained. 
Here, we now applied the adiabatic approximation and also switched to a time-dependent description. 
This equation is inserted into Eq.~(\ref{eqJ1}) on our way of closing our dynamical equations.

Based on Eqs.~(\ref{nbody}), (\ref{rhoeq_equilibrium_prev}), and (\ref{Boltzmann}), i.e., again applying the adiabatic approximation, we find two further helpful relations,
namely
\begin{align}
k_\mathrm{B} &T \, \nabla_{\vect{r}'}\rh{2,0}( \vect{X},\vect{X}',t) \nonumber
+\rh{2,0}( \vect{X},\vect{X}',t)\nabla_{\vect{r}'}u^\mathrm{AA}(\revi{|\vect{r}-\vect{r}'|}) \nonumber\\&
+ \! \!\int \! \! \mathrm{d}\vect{X}'' \rh{2,1}(\vect{X},\vect{X}',\vect{X}'', t)\nabla_{\vect{r}'} u^\mathrm{AB}(\revi{|\vect{r}'-\vect{r}''|}) \nonumber\\ &
+ \! \!\int \! \! \mathrm{d}\vect{X}'' \rh{3,0}(\vect{X},\vect{X}',\vect{X}'', t)\nabla_{\vect{r}'} u^\mathrm{AA}(\revi{|\vect{r}'-\vect{r}''|}) \nonumber \\
= \; &k_\mathrm{B} T \, \rh{2,0}(\vect{X},\vect{X}',t) \nabla_{\vect{r}'} \ln \left(\lambda_\mathrm{A}^3 \, \rho_\mathrm{A}(\vect{X}',t)\right) \nonumber \\&+ 
\rh{2,0}(\vect{X},\vect{X}',t) \nabla_{\vect{r}'} \frac{\delta {\mathcal{F}}_{\!\!\mathrm{exc}}}{\delta \rho_\mathrm{A}(\vect{X}',t)} 
\end{align}
and
\begin{align}
 k_\mathrm{B} &T \, \nabla_{\vect{r}'}\rh{1,1}( \vect{X},\vect{X}',t) 
+\rh{1,1}( \vect{X},\vect{X}',t)\nabla_{\vect{r}'}u^\mathrm{AB}(\revi{|\vect{r}-\vect{r}'|}) \nonumber\\&
+ \! \! \int \! \! \mathrm{d}\vect{X}'' \rh{1,2}(\vect{X},\vect{X}',\vect{X}'', t)\nabla_{\vect{r}'} u^\mathrm{BB}(\revi{|\vect{r}'-\vect{r}''|}) \nonumber\\&
+ \! \! \int \! \! \mathrm{d}\vect{X}'' \rh{2,1}(\vect{X},\vect{X}'',\vect{X}', t)\nabla_{\vect{r}'} u^\mathrm{AB}(\revi{|\vect{r}'-\vect{r}''|}) \nonumber \\
= \; &k_\mathrm{B} T \, \rh{1,1}(\vect{X},\vect{X}',t) \nabla_{\vect{r}'} \ln \left(\lambda_\mathrm{B}^3 \, \rho_\mathrm{B}(\vect{X}',t)\right) \nonumber \\&+ 
\rh{1,1}(\vect{X},\vect{X}',t) \nabla_{\vect{r}'} \frac{\delta {\mathcal{F}}_{\!\!\mathrm{exc}}}{\delta \rho_\mathrm{B}(\vect{X}',t)} \, .
\label{mixed_ybg}
\end{align}
\allowdisplaybreaks
Analogues for species~B follow after applying to Eqs.~(\ref{ybg_a})--(\ref{mixed_ybg}) the replacements listed below Eq.~(\ref{eqK6}).

Inserting the above relations into Eqs.~(\ref{eqJ1}), (\ref{eqKA1}), (\ref{eqKA4}), (\ref{eqK1}), and (\ref{eqK4}) yields
\allowdisplaybreaks[0]
\begin{align}
\curr{\mathrm{A}}^\mathrm{tt}
=&
{}-\mu^\mathrm{t}_\mathrm{A}\bigg( k_\mathrm{B} T \, \nabla_{\mathbf r}\rho_\mathrm{A}(\vect{X},t)
+\rho_\mathrm{A}(\vect{X},t)\,\nabla_{\mathbf r}\,u^\mathrm{A}_\mathrm{ext}(\mathbf r)
\notag\\
&+\rho_\mathrm{A}(\vect{X},t)\nabla_{\mathbf r}\frac{\delta {\mathcal F}_{\! \! \mathrm{exc}}}{\delta\rho_\mathrm{A}(\vect{X},t)}\bigg),
\label{eqJ1r} \\ 
\currk{\mathrm{AA}}^\mathrm{tt}
=
&-\! \! \int \! \! \mathrm{d}\vect{X}' \, \rh{2,0}(\vect{X},\vect{X}',t)\ 
\,{\vgr{\mu}^{\mathrm{tt},\mathrm{AA}}_{\vect{r}, \vect{r}'}} \cdot\vect{j}_\mathrm{A}(\vect{X}',t),
\label{eqKA1r} \\
\currk{\mathrm{AA}}^\mathrm{rt}
=
&-\! \! \int \! \! \mathrm{d}\vect{X}' \, \rh{2,0}(\vect{X},\vect{X}',t)\ 
\,{\vgr{\mu}^{\mathrm{rt},\mathrm{AA}}_{\vect{r}, \vect{r}'}} \,\, \vect{j}_\mathrm{A}(\vect{X}',t),
\label{eqKA4r} \\
\currk{\mathrm{AB}}^\mathrm{tt}
=
&-\! \! \int \! \! \mathrm{d}\vect{X}' \, \rh{1,1}(\vect{X},\vect{X}',t)\ 
\,{\vgr{\mu}^{\mathrm{tt},\mathrm{AB}}_{\vect{r}, \vect{r}'}} \cdot \vect{j}_\mathrm{B}(\vect{X}',t),
\label{eqK1r} \\
\currk{\mathrm{AB}}^\mathrm{rt}
=
&-\! \! \int \! \! \mathrm{d}\vect{X}' \, \rh{1,1}(\vect{X},\vect{X}',t)\ 
\,{\vgr{\mu}^{\mathrm{rt},\mathrm{AB}}_{\vect{r}, \vect{r}'}} \,\, \vect{j}_\mathrm{B}(\vect{X}',t),
\label{eqK4r}
\end{align}
respectively, where we defined the vector fields
\begin{align}
 \vect{j}_\nu(\vect{X}',t) = &\,k_\mathrm{B} T \, \nabla_{\vect{r}'} \ln\left(\lambda_\nu^3 \, \rho_\nu(\vect{X}',t)\right)
\notag\\&
 + \nabla_{\vect{r}'} \!\left( u^\nu_\mathrm{ext}(\vect{r}')+\frac{\delta {\mathcal F}_{\!\!\mathrm{exc}}}{\delta\rho_\nu(\vect{X}',t)} \right).
\label{eq:j_nu}
\end{align}

This way, the two-swimmer density in Eq.~(\ref{eqJ1}) and all three-swimmer densities have been eliminated.
Again, analogous relations apply to the dynamical equation for~$\rho_\mathrm{B}(\vect{x},t)$ and are obtained by considering the replacements introduced below Eq.~(\ref{eqK6}).

Still, the remaining two-swimmer densities in the $\currk{\, \cdot}^{\cdot \cdot}$ current densities must be addressed.
For this purpose, as in a previous work,\cite{hoell2017dynamical} we employ the Onsager-type \cite{onsager1949effects} approximations
\begin{align}
 \rh{2,0}(\vect{X},\vect{X}',t) = &
 \rho_\mathrm{A}(\vect{X},t) \, \rho_\mathrm{A}(\vect{X}',t) \notag \\ &\times
 \exp\left(-\beta u^\mathrm{AA}\left(\revi{|\vect{r}-\vect{r}'|}\right)\right)\!, 
\label{Ons20} \\
 \rh{1,1}(\vect{X},\vect{X}',t) = &
 \rho_\mathrm{A}(\vect{X},t) \, \rho_\mathrm{B}(\vect{X}',t)  \notag \\ &\times
 \exp\left(-\beta u^\mathrm{AB}\left(\revi{|\vect{r}-\vect{r}'|}\right)\right)\!, 
\label{Ons11} \\
 \rh{0,2}(\vect{X},\vect{X}',t) = &
 \rho_\mathrm{B}(\vect{X},t) \, \rho_\mathrm{B}(\vect{X}',t)  \notag \\ &\times
 \exp\left(-\beta u^\mathrm{BB}\left(\revi{|\vect{r}-\vect{r}'|}\right)\right)\!.
\label{Ons02}
\end{align}
Here, for $|\vect{r}-\vect{r}'|$ smaller than the sum of the radii of the involved swimmer bodies, we furthermore set the pair densities to zero to avoid the otherwise-appearing unphysical hydrodynamic divergences.
Strictly speaking, this leads to a discontinuity, but typically the jump is vanishingly small, e.g., $\exp(-5 \exp(-1/16))\approx 0.009 \ll 1$ for $\epsilon_0^{\cdot \cdot} = 5 k_\mathrm{B} T$ and $a_\cdot=\sigma_\cdot /4$, see Eq.~(\ref{interaction_potential}). This order of magnitude is sufficiently low to treat the function as basically ``smooth'' in the numerical evaluation.

Equations~(\ref{Ons20})--(\ref{Ons02}) implicitly assume $g_{\mu \nu}(\vect{X},\vect{X}',t)\approx\exp\left(-\beta u^{\mu \nu}(|\vect{r}-\vect{r}'|)\right)$ for the pair distribution functions, with $\mu,\nu \in \{ \mathrm{A,B} \}$.
Using these relations is exact for passive equilibrium systems in the low-density limit,\cite{hansen1990theory} as the expressions are based on the assumption that the two involved particles interact only with each other (and with no third particles).
Adapting these relations to describe semi-dilute active suspensions thus constitutes a reasonable first-order approximation beyond assuming a constant pair distribution function.
More generally, one could at this point also insert another reasonable approximation for the pair distribution function.

Similarly, our (pairwise) treatment of hydrodynamic interactions between the swimmers, see Eqs.~(\ref{mobility_og})--(\ref{defminus}), requires sufficiently large distances between the swimmer bodies. 
First, this is ensured by the steric interaction between the swimmers when half of its effective range, i.e., $\sigma_{\mu\nu}/2$ in Eq.~(\ref{interaction_potential}), is larger than~$a_\kappa$, $\alpha_\kappa L_\kappa$, and~$(1-\alpha_\kappa) L_\kappa$, with~$\mu,\nu\in\{\mathrm{A,B}\}$ and~$\kappa \in \{\mu,\nu\}$.
The larger the mean distances are between the swimmers, the higher the accuracy of our description of hydrodynamic interactions will be.
Together with the assumptions involved in Eqs.~(\ref{Ons20})--(\ref{Ons02}), we thus expect our DDFT for multi-species systems of microswimmers to perform best for \mbox{(semi-)dilute} suspensions of swimmers, within which our steric interaction potentials maintain a significant distance between the swimmer bodies, even when they are heading for collisions.

Finally, the excess functional ${\mathcal F}_{\! \! \mathrm{exc}}$ involving the effective steric interactions between the swimmers needs to be specified.
As appropriate for GEM potentials,\cite{archer2014solidification}
we from now on use a mean-field approximation, here for our case of binary mixtures, reading
\begin{equation}
 {\mathcal F}_{\! \! \mathrm{exc}} = \frac{1}{2}
 \! \int \! \! \mathrm{d}\vect{X} \! \! \int \! \! \mathrm{d}\vect{X}' \,
\rho_\mu (\vect{X},t) \, \rho_\nu(\vect{X}', t) \, \,
 u^{\mu\nu}\!\left(\revi{|\vect{r}-\vect{r}'|}\right),
\label{Fexc}
\end{equation}
with $\mu,\nu \in \{\mathrm{A,B}\}$ and summing over repeated indices.
In this way, our set of coupled dynamical equations for $\rho_\mathrm{A}(\vect{X},t)$ and $\rho_\mathrm{B}(\vect{X},t)$ is closed.
We remark that, along the same lines, a theory for more than two different species can be derived as well, leading to a correspondingly further increased number of terms.
Here, we continue by applying the above theory to concrete example situations in Sec.~\ref{sec:results}.

\section{Applications}
\label{sec:results}

\allowdisplaybreaks[0]

In this section, the DDFT derived in Sec.~\ref{sec:ddft} is applied to several illustrative cases.
Specifically, for simplicity, these will be setups in which the positions and orientations of the swimmers are constricted to the $xy$-plane.
Still, a surrounding bulk fluid is considered with the planar swimmer ensemble embedded therein, allowing for three-dimensional fluid flows.
Possible methods to experimentally realize this situation could be the confinement of microswimmers to the interface between two immiscible fluids of identical viscosity~$\eta$, or the use of optical trapping fields.

In such a setup, the orientation of a swimmer is described by a single angle $\phi$ (measured from the $x$-axis) via 
$\uvec{n}=(\cos\phi, \sin\phi)$. 
The orientational gradient operator then reduces to 
$\uvec{n} \times \nabla_{\uvec{n}}=\uvec{z} \partial_\phi $, where 
$\uvec{z}$ is the oriented Cartesian unit vector pointing (upwards) out of the $xy$-plane.
Furthermore, the phase-space coordinate $\vect{X}$ in this situation becomes $\vect{X}=\{x,y,\phi\}$.

\begin{figure*}
 \includegraphics[trim=0 10 0 10, clip]{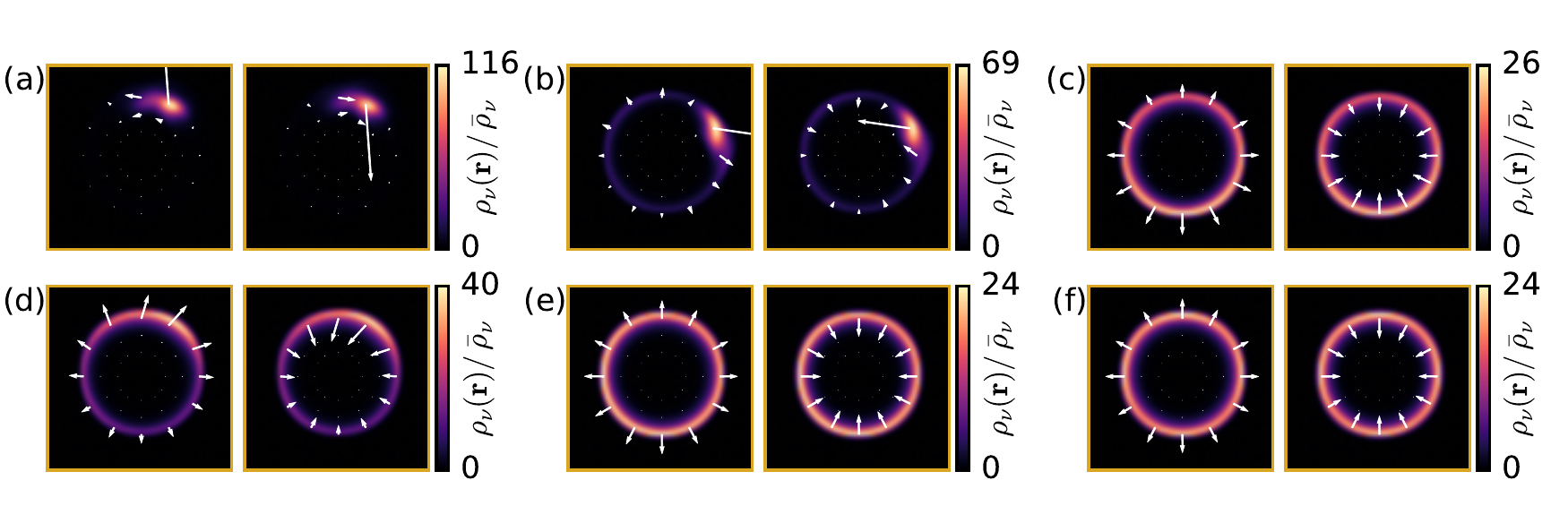} 
\caption{
Steady-state density distribution for binary mixtures of pusher~(A) and puller~(B) microswimmers in an external trapping potential, see Eq.~(\ref{trap}), for varying overall densities $\bar{\rho}_\mathrm{A}$ (pushers) and $\bar{\rho}_\mathrm{B}$ (pullers).
All other parameters are held constant at $a_\mathrm{A}=a_\mathrm{B}=0.25 \, \sigma$, $L_\mathrm{A}=L_\mathrm{B}=0.75 \, \sigma$, $\alpha_\mathrm{A}=\alpha_\mathrm{B}=0.4$, $V_0^\mathrm{A}=V_0^\mathrm{B}=0.5 \, k_\mathrm{B} T$,
$\epsilon_0^\mathrm{A}=\epsilon_0^\mathrm{B}=10 \, k_\mathrm{B} T$, and~\mbox{$f_\mathrm{A}={}-f_\mathrm{B}=600 \, k_\mathrm{B} T / \sigma$}, with $ \sigma_\mathrm{A} =\sigma_\mathrm{B} \equiv \sigma$. 
The simulation box is of size~$18 \, \sigma \times 18 \, \sigma$ (only the inner $12\, \sigma\times12\, \sigma$ are on display) and the numerical evaluations were performed on ($80\times80\times16$)-grids. 
Each pair of plots shows on the left-hand side the results for species~A (pushers) and on the right-hand side the corresponding distribution for species~B (pullers). 
In each plot, the color encodes the (reduced) spatial density profile~$\rho_\nu (\vect{r},t) / \bar{\rho}_\nu $ 
(reduced by the average density~$\bar{\rho}_\nu$),
with brighter color corresponding to higher density,
and white arrows indicate the orientational vector field~$\left<\uvec{n}\right>_\nu(\vect{r},t)$, as defined in Eqs.~(\ref{dens_profile}) and~(\ref{or_vf}), respectively.
The overall densities~$(\bar{\rho}_\mathrm{A},\bar{\rho}_\mathrm{B})$ are given (in units of~$\sigma^{-2}$) by (a)~$(0.0123,0.00617)$, (b)~$(0.0123,0.00926)$, (c)~$(0.0123,0.0123)$, (d)~$(0.00926,0.00617)$, (e)~$(0.00926,0.00926)$, and (f)~$(0.00926,0.0123$). 
The systems in~(a),~(b), and~(d) do not reach steady states in a strict sense, as the spot formation there is unstable against (spontaneous) movement of the density profile along the rim of the trap.
}
\label{fig:trap_results}
\end{figure*}

The numerical solution of the coupled set of partial differential equations derived in Sec.~\ref{sec:ddft} is then performed on an equidistant $N_x \times N_y \times N_\phi$ grid using the finite-volume-method solver \emph{FiPy}\cite{Guyer_2009_CiSE}.
Formally, numerical periodic boundary conditions are imposed on all coordinates~$x$, $y$, and~$\phi$, but hydrodynamic and steric interactions are cut at a distance chosen such that no (unphysical) interactions across the boundaries occur. 
As nevertheless all physical interactions inside the system should, of course, be accounted for, we further always set the length of the simulation box in both spatial directions to at least twice the largest relevant interparticle distance.

Since the orientation-dependent densities $\rho_\nu(\vect{X},t)$ at time $t$ are still a function of~$x$, $y$, and~$\phi$,
they cannot be easily plotted even for our planar configurations.
For displaying our results, we thus further define the (orientation-integrated) spatial swimmer densities
\begin{equation}
 \rho_\nu(\vect{r},t) = \!\! \int\limits_0^{2\pi} \!\! \mathrm{d} \phi \, \, \rho_\nu(\vect{X},t)
\label{dens_profile}
\end{equation}
and the orientational vector fields
\begin{equation}
 \left< \uvec{n} \right>_\nu (\vect{r},t) = \!\! \int\limits_0^{2\pi}  \!\! \mathrm{d} \phi \, \, \uvec{n}(\phi) \, \rho_\nu(\vect{X},t),
\label{or_vf}
\end{equation}
where $\nu \in \{\mathrm{A,B}\}$.
Moreover, the overall (average) one-species densities are described by $\bar{\rho}_\nu = A^{-1} \!\! \int_A \!\mathrm{d} \vect{r} \, \rho_\nu(\vect{r},t)$, where $A$ is the area of the regarded system.

\subsection{Trapped binary swimmer system}
\label{subsec:results_trap}

While restricting the binary microswimmer configuration to two spatial dimensions as detailed above, we now additionally introduce radially-symmetric quartic trapping potentials given by 
\begin{equation}
 u^\nu_\mathrm{ext}(\vect{r}) = V^\nu_0 \left( \frac{r}{\sigma} \right)^4,
\label{trap}
\end{equation}
with potential strengths $V^\nu_0$, distance $r=|\vect{r}|$ to the center of the trap, and $\nu=\mathrm{A, B}$.
As in previous works,\cite{menzel2016dynamical,hoell2017dynamical} we use a quartic potential --- instead of, e.g., a harmonic one ($\propto r^2$) --- to observe more pronounced differences between activity-induced off-center density distributions (see below) and center-heavy equilibrium distributions for passive particles.
Previously reported results for harmonic traps\cite{nash2010run, hennes2014self} showed qualitative agreement with our results for a quartic potential.\cite{menzel2016dynamical,hoell2017dynamical}
For simplicity, we furthermore from now on assume that all species-related parameters are the same for both species, except for \mbox{$f_\mathrm{A}=-f_\mathrm{B}>0$}.
Thus, species~A is formed by pushers and species~B represents pullers (of the same strength). 

In analogous one-component suspensions,\cite{menzel2016dynamical,hoell2017dynamical} 
without any active drive, the external potential leads to center-heavy distributions following standard equilibrium statistics.
When the active drive is switched on in the one-component systems, but hydrodynamic interactions are still neglected, the self-propelled particles start forming a radially symmetric high-density ring, along which the outward self-propulsion is balanced by the restoring trapping force.\cite{menzel2015focusing,yan2015force,menzel2016dynamical}
With hydrodynamic interactions incorporated, this ring of microswimmers can become unstable against collapsing to one spot on this ring, which is induced by the hydrodynamic coupling through the resulting fluid flows.\cite{nash2010run,hennes2014self,menzel2016dynamical,hoell2017dynamical} 
In parts of the parameter space, pushers and pullers were observed to behave quite differently, with pushers showing a significantly more pronounced destabilization of the high-density ring and formation of a high-density spot, while pullers showed a much weaker density variation along the ring.\cite{hoell2017dynamical}

We are now interested in pusher--puller mixtures. 
There is a crucial competition between hydrodynamic effects resulting from the external potential acting on the swimmer bodies and from the actively introduced forces exerted by the microswimmers themselves.
We here concentrate on a parameter range for which the hydrodynamic interactions  induced by the self-propulsion mechanism dominate those induced by the external potential force.
Concerning our current densities, we thus always check that
$|\currk{\, \cdot \cdot}^{\mathrm{ra}}|>|\currk{\, \cdot \cdot}^{\mathrm{rt}}|$, see, e.g., Eqs.~(\ref{eqK4}) and (\ref{eqK6}), for our chosen parameters.

Numerical results for (steady-state) distributions of pusher--puller mixtures are shown in Fig.~\ref{fig:trap_results}, for varying overall densities of the two species.
In strong contrast to the corresponding one-component systems, for which the (steady-state) distributions strongly differed between pure pusher and pure puller systems,\cite{hoell2017dynamical} we here frequently observe the same qualitative behavior when both species are present simultaneously.
For instance, in Fig.~\ref{fig:trap_results}~(a), pushers transfer their ``spot-forming'' tendency onto the pullers, which in the absence of the pushers would show a ring-like arrangement instead of the spot.
However, the plots in Fig.~\ref{fig:trap_results} indicate the rough relation~$\rho_\mathrm{A}(\vect{r},\uvec{n},t) / \bar{\rho}_\mathrm{A} \approx \rho_\mathrm{B}(\vect{r},{}-\uvec{n},t) / \bar{\rho}_\mathrm{B}$.
Choosing, e.g., $|f_\mathrm{A}|\neq|f_\mathrm{B}|$, this approximate relation breaks down as the two species aggregate at different distances from the origin, but for sufficiently small deviations, we still observe a qualitatively similar collective behavior for both species.

In Fig.~\ref{fig:trap_results}, the overall density $\bar{\rho}_\mathrm{B}$ of pullers increases from left to right, 
while the overall density $\bar{\rho}_\mathrm{A}$ for pushers decreases from the top row to the bottom row.
We observe clear spot formation in Fig.~\ref{fig:trap_results}~(a) and~(b), while~(c) shows less-pronounced instabilities of the high-density ring.
Thus, we may conclude, that the dominating species imposes its behavior onto the other species.

For Fig.~\ref{fig:trap_results}~(c) and~(e), where~$\bar{\rho}_\mathrm{A}=\bar{\rho}_\mathrm{B}$ and therefore~$\rho_\mathrm{A}(\vect{r},\uvec{n},t)\approx \rho_\mathrm{B}(\vect{r},-\uvec{n},t)$ holds, the probability currents associated with the rotation due to the active forces approximately cancel each other by symmetry, e.g.,~$\currk{\mathrm{AA}}^\mathrm{ra}\approx{}-\currk{\mathrm{AB}}^\mathrm{ra}$, so that only the currents $\currk{\, \cdot \cdot}^\mathrm{rt}$ can lead to spot formation.
The latter starts to outperform the rotational diffusion for the case depicted in Fig.~\ref{fig:trap_results}~(c), but not for the lower overall densities in Fig.~\ref{fig:trap_results}~(e).
The instability of the ring here seems to be a question of high-enough overall density because, e.g.,~$|\currk{\mathrm{AA}}^\mathrm{rt}| \propto \bar{\rho}_\mathrm{A}^2$ and~$ |\curr{\mathrm{A}}^\mathrm{rr}|\propto \bar{\rho}_\mathrm{A}$.

The bottom row of Fig.~\ref{fig:trap_results} shows the corresponding density distributions for a smaller $\bar{\rho}_\mathrm{A}$.
Thus, a decreased density of pushers leads to an increased stability of the high-density ring against aggregation in one spot. 
When (significantly) more pullers than pushers are in the system, as in Fig.~\ref{fig:trap_results}~(f), they dominate the overall behavior and re-stabilize the high-density ring. 
 
In summary, the majority species seems to dominate the overall behavior of the system.
A similar conclusion has recently been drawn for the unconfined motion in pusher--puller mixtures,\cite{pessot2018} which we will treat as the next example using our theoretical approach.

At this point, we include a short remark on the performance of our theory. 
We can remove the second species from our DDFT equations derived in Sec.~\ref{sec:ddft} by setting \mbox{$\rho_\mathrm{B}(\vect{X},t)\equiv0$}.
Then, the present set of equations reduces to the previous DDFT for monodisperse microswimmers.\cite{menzel2016dynamical}
In that case, likewise, the statistical theory was evaluated by exposing the system of swimmers to a radial external trapping potential, in analogy to the above consideration for a pusher--puller mixture.
There, hydrodynamic interactions lead to the formation a high-density spot of aligned swimmers as well, resulting in overall flow fields.\cite{menzel2016dynamical, hoell2017dynamical} 
This ``hydrodynamic fluid pump'' had previously been reported in particle-based computer simulations,\cite{nash2010run,hennes2014self} using different swimmer models.
Thus, a qualitative comparison shows that our DDFT reproduces corresponding general phenomena.
Adding another microswimmer species to the same framework, we expect a similarly successful performance of the present theory.
Direct quantitative comparison could be carried out in the future by implementing a suitable particle picture into many-swimmer computer simulations including hydrodynamic interactions and thermal fluctuations, e.g., via multiparticle collision dynamics\cite{malevanets1999mesoscopic,kapral2008multiparticle,gompper2009multi,zottl2012nonlinear,zottl2014hydrodynamics,ruhle2018gravity} / stochastic rotation dynamics.\cite{ihle2001stochastic,downton2009simulation}
Then, also higher swimmer densities could be addressed numerically.
Another way to explicitly take into account the induced hydrodynamic fluid flows in computer simulations could be Lattice-Boltzmann methods.\cite{mcnamara1988use,nash2010run,alarcon2013spontaneous,kuron2019lattice,bardfalvy2019particle}

\subsection{Emergence of polar orientational order and collective motion in pusher--puller mixtures}

In the absence of the spherical trapping potential considered in Sec.~\ref{subsec:results_trap},
previous particle-based computer simulations of planar arrangements of microswimmers with periodic boundary conditions and using the same swimmer model have identified a tendency of puller microswimmers to develop (global) collective polar orientational order.\cite{pessot2018} 
Related observations were made in simulations of analogous three-dimensional configurations of squirmer microswimmers.\cite{alarcon2013spontaneous}
Such order in the swimmer orientations naturally leads to collective motion, maintaining a common average propulsion direction.
Moreover, we have performed a corresponding linear stability analysis of our DDFT for planar pure (one-species) pusher or puller systems, with spontaneous ordering identified beyond a threshold active drive for pullers,\cite{hoell2018particle} in contrast to pushers.
We now address the corresponding two-species situation.
In related computer simulations for mixtures of pushers and pullers using the same swimmer model,\cite{pessot2018} it was found that collective orientational order only develops if the fraction of pushers is sufficiently small.
As we demonstrate, our DDFT reproduces these results and leads to a more quantitative insight.

For this purpose, the external potential in our planar arrangement is now set to~$u_\mathrm{ext}(\vect{r})\equiv 0$.
For simplicity, we assume that the one-swimmer densities are spatially homogeneous, i.e., $\rho_\nu (\vect{X},t) = \rho_\nu(\phi,t)/A$, with $A$ denoting the area (considered to be large) of the periodic plane containing the swimmers and $\nu \in \{\mathrm{A,B}\}$. 
Then, integrating Eq.~(\ref{BBGKY1}) over all positions~$\vect{r}$ in the periodic box leads to 
\begin{align}
 \!\!\frac{\partial\rho_\mathrm{A}(\phi,t)}{\partial t} = 
&{}-\uvec{z} \cdot \! \! \int \!\! \mathrm{d}\vect{r} \, \frac{\partial}{\partial \phi} \Big(\curr{\mathrm{A}}^\mathrm{rr} 
	+ \!\!\!\!\sum\limits_{\nu =\mathrm{A,B}} \! \!\!\! \left(\currk{\mathrm{A}\nu}^\mathrm{rt}\!+\! \currk{\mathrm{A}\nu}^\mathrm{ra}\right)\Big),
\label{order_BBGKY}
\end{align}
with the probability current densities defined in Eqs.~(\ref{eqJ1})--(\ref{eqK6}).
Following Ref.~\onlinecite{hoell2018particle}, the current densities $\currk{\cdot \cdot}^\mathrm{rt}$ are neglected for sufficiently dilute suspensions,
as all the contained non-vanishing terms scale with three-swimmer densities.
Thus, Eq.~(\ref{order_BBGKY}) reduces to
\begin{align}
 &\frac{\partial\rho_\mathrm{A}(\phi,t)}{\partial t} =  \, k_\mathrm{B} T \, \mu^{\mathrm{r,A}} \, \partial_\phi^2 \rho_\mathrm{A}(\phi,t) \notag \\
&{}- f_\mathrm{A}  \, \partial_\phi \! \int \! \! \mathrm{d} \vect{r} \! \int \! \! \mathrm{d} \vect{X}' \, \uvec{z} \cdot \left( \vgr{\Lambda}^{\mathrm{rt},\mathrm{AA}}_{\vect{r},\vect{X}'} \, \uvec{n}'\right) \rh{2,0}(\vect{X},\vect{X}',t) \notag \\
&{}- f_\mathrm{B} \, \partial_\phi \! \int \! \! \mathrm{d} \vect{r} \! \int \! \! \mathrm{d} \vect{X}' \, \uvec{z} \cdot \left(\vgr{\Lambda}^{\mathrm{rt},\mathrm{AB}}_{\vect{r},\vect{X}'} \, \uvec{n}' \right) \rh{1,1}(\vect{X},\vect{X}',t).
\label{order1}
\end{align}
Here, the two-swimmer densities are related to the pair distribution functions via
\begin{align}
 \rh{2,0}(\vect{X},\vect{X}',t) = \frac{\rho_\mathrm{A}(\phi,t) \, \rho_\mathrm{A}(\phi',t) \, g_\mathrm{AA}(\vect{X},\vect{X}',t)}{A^2} \, , \\
 \rh{1,1}(\vect{X},\vect{X}',t) = \frac{\rho_\mathrm{A}(\phi,t) \, \rho_\mathrm{B}(\phi',t) \, g_\mathrm{AB}(\vect{X},\vect{X}',t)}{A^2} \, .
\end{align}
Thus, Eq.~(\ref{order1}) becomes
\begin{align}
 &\frac{\partial\rho_\mathrm{A}(\phi,t)}{\partial t} = \, k_\mathrm{B} T \, \mu^{\mathrm{r,A}} \, \partial_\phi^2 \rho_\mathrm{A}(\phi,t) \notag \\
&{}- f_\mathrm{A} \, \partial_\phi \left[ \rho_\mathrm{A}(\phi,t) \! \int \! \! \mathrm{d} \phi' \, \rho_\mathrm{A}(\phi',t) \, G_\mathrm{AA}(\phi-\phi',t) \right] \notag \\
&{}- f_\mathrm{B} \, \partial_\phi \left[ \rho_\mathrm{A}(\phi,t) \! \int \! \! \mathrm{d} \phi' \, \rho_\mathrm{B}(\phi',t) \, G_\mathrm{AB}(\phi-\phi',t) \right],
\label{order2}
\end{align}
where the hydrodynamic interactions are comprised in the coupling functions
\begin{align}
 G_{\mu \nu}(\phi-\phi',t) := \! \! \int \! \! \mathrm{d} \vect{r} \! \! \int \! \! \mathrm{d} \vect{r}' \, \frac{\uvec{z} \cdot \left( \vgr{\Lambda}^{\mathrm{rt},\mu \nu}_{\vect{r},\vect{X}'} \, \uvec{n}'\right) g_{\mu \nu}(\vect{X},\vect{X}',t)}{A^2} \, ,
\label{Gmunu}
\end{align}
with~$\mu, \nu \in \{\mathrm{A,B}\}$.
An analogous dynamical equation for species~B is obtained by replacing 
$\mathrm{A}\to\mathrm{B}$ and~$\mathrm{B}\to\mathrm{A}$.
In the following,  species~A again represents pushers, and species~B pullers.

To allow for further analytical treatment, we include additional simplifying assumptions.
Considering systems in which all active agents propel with the same amplitude of the active drive and further are identical in all other microscopic parameters, 
the coupling and pair distribution functions, see Eq.~(\ref{Gmunu}), were determined in Ref.~\onlinecite{hoell2018particle} by a modified Percus test-particle method.
For this purpose, hydrodynamic interactions were neglected and only the interplay of self-propulsion and steric interactions was evaluated.
As a result, we had extracted and approximated the basic functional form as \cite{hoell2018particle}
\begin{equation}
G_{\mu \nu}(\phi-\phi')= \tilde{C}_{\mu \nu} \sin(\phi-\phi'),
\label{G_approx}
\end{equation}
where $\tilde{C}_\mathrm{AA}=\tilde{C}_\mathrm{BB}=\tilde{C}/A>0$ is positive for same-species coupling, and $\tilde{C}_\mathrm{AB}=\tilde{C}_\mathrm{BA}={}-\tilde{C}/A$.
This distinction follows from the fact of our puller microswimmers propelling into the direction of ${}-\uvec{n}$ and~/~or ${}-\uvec{n}'$, see Fig.~\ref{fig:model}.
Since $\phi$ and $\phi'$ parameterize the orientations of $\uvec{n}$ and $\uvec{n}'$, respectively, the swimming direction of a puller is shifted by an additional angle $\pi$ relatively to $\phi$ and~/~or $\phi'$.
If only one of the angles $\phi$ and $\phi'$ refers to a puller, the additional shift of $\phi-\phi'$ by $\pi$ requires a minus sign in the prefactor of $\sin(\phi-\phi')$ in Eq.~(\ref{G_approx}).

The value of~$\tilde{C}>0$ generally depends on the overall density and the microscopic parameters. 
(Some further positive constant parameters are here incorporated by the coefficients $\tilde{C}$ when compared to the amplitude $C$ in Ref.~\onlinecite{hoell2018particle}.)
Since a similarly simple analytically treatable expression is still missing for hydrodynamic interactions included on the level of pair distribution functions, we use Eq.~(\ref{G_approx}) as an input for our further calculations.

We assume that, if collective order arises, there is only one common direction of polar ordering, i.e., that in this case species~A and~B collectively propel along a common direction.
This assumption is motivated by previous simulation results.\cite{pessot2018}
We now test the linear stability of the uniform distributions~$\rho_\nu(\phi,t)\equiv N_\nu / (2 \pi)$ against the emergence of collective orientational ordering.
To this end, the ansatz~\mbox{$\rho_\mathrm{A}(\phi,t) = N_\mathrm{A} / (2 \pi) + \epsilon_\mathrm{A}(t) \cos(\phi-\phi_0)$} and \mbox{$\rho_\mathrm{B}(\phi,t) = N_\mathrm{B} / (2 \pi) + \epsilon_\mathrm{B}(t) \cos(\phi-\phi_0+\pi)$} ,
with~an arbitrary angle~$\phi_0$ and~$|\epsilon_\nu(t)| \ll N_\nu$ for $\nu\in \{\mathrm{A,B}\}$,
is inserted into Eq.~(\ref{order2}) and the equivalent equation for species~B.
This leads to the coupled ordinary differential equations
\begin{align}
\frac{\mathrm{d}}{\mathrm{d} \, t}
\left[ \hspace{-0.8 ex}
\begin{array}{c}
 \epsilon_\mathrm{A}(t) \\[0.1cm]
 \epsilon_\mathrm{B}(t)
\end{array}\hspace{-0.8 ex}
\right]
=    
\vect{M} \cdot 
\left[ \hspace{-0.8 ex}
\begin{array}{c}
 \epsilon_\mathrm{A}(t) \\[0.1cm]
 \epsilon_\mathrm{B}(t)
\end{array} \hspace{-0.8 ex}
\right],
\end{align}
with the coefficient matrix
\begin{align}
	\vect{M}={}-
\left[ 
\hspace{-0.8 ex}
	\begin{array}{cc}
    k_\mathrm{B} T  \mu^{\mathrm{r,A}} +
	 m_\mathrm{AA} & m_{\mathrm{AB}}\\[0.1cm]
     m_{\mathrm{BA}} & k_\mathrm{B} T \mu^{\mathrm{r,B}} 
	+ m_{\mathrm{BB}}  
    \end{array} 
\hspace{-0.8 ex}
\right],
\end{align}
where $m_{\mu \nu}:=   N_\mu f_\nu \tilde{C} /(2 A) $.

We recall that species A~(pushers) and species~B (pullers) are considered to have the same amplitude of their active drive, i.e.,~\mbox{$f_\mathrm{A} = -f_\mathrm{B} > 0$}.
Additionally, we keep \mbox{$N_\mathrm{A}+N_\mathrm{B}=N$} constant, i.e., only the ratio of pushers to pullers is varied.
Moreover, all other parameters are assumed to be identical for the two species.
Then, the eigenvalues of $\vect{M}$ are determined as 
$\left(\!{}-k_\mathrm{B} T  \mu^{\mathrm{r,A}}, {}-k_\mathrm{B} T  \mu^{\mathrm{r,A}} + f_\mathrm{A} \tilde{C} N ( \chi_\mathrm{B} -1 / 2) / A \right)$.
Here, the first eigenvalue is always negative, but the second one becomes positive if
\begin{equation}
 k_\mathrm{B} T  \mu^{\mathrm{r,A}} < f_\mathrm{A} \tilde{C} \, \frac{N}{A} \, (\chi_\mathrm{B}-1 / 2),
\label{order_criterion}
\end{equation}
with~$\chi_\mathrm{B}:=N_\mathrm{B} / N$ denoting the fraction of pullers.
The corresponding eigenvector is ($N_\mathrm{A},N_\mathrm{B})$. 

Our system can thus be linearly unstable against polar orientational ordering only if the right-hand side of Eq.~(\ref{order_criterion}) is positive. 
Since $f_\mathrm{A}>0$, this implies that the pullers must outnumber the pushers ($\chi_\mathrm{B}>1/2$).
If this condition is satisfied, the active drive additionally needs to be strong enough, i.e., Eq.~(\ref{order_criterion}) sets a threshold strength for $f_\mathrm{A}=-f_\mathrm{B}$.
Particularly, the effect of the active drive and the hydrodynamic interactions need to outperform rotational diffusion.
Furthermore, as indicated by the corresponding eigenvector $(N_\mathrm{A},N_\mathrm{B})$, if orientational order arises, it does so simultaneously for both species.

Our results roughly agree with those in the previous simulation study.\cite{pessot2018}
We stress that our theory only tests linear instability with respect to polar orientational ordering and that the above approximations were involved.
In particular, the influence of hydrodynamic interactions on the pair distribution function was neglected.
To address this question, possibly the results of particle-based computer simulations could be used as an input to the theory in the future.\cite{pessot2018,schwarzendahl2019hydrodynamic}
Since our previous theoretical analysis for single-species systems indicated polar orientational ordering for puller suspensions but not for pushers,\cite{hoell2018particle}
we again find that the majority species imposes its behavior onto the minority species, as observed already for the confined (trapped) mixtures in Sec.~\ref{subsec:results_trap}.

\subsection{Shear cell}

As a third example, we now address a planar circular configuration which effectively represents a shear cell. 
We compose this shear cell of passive colloidal particles forming an effective circular rim and active microswimmers trapped inside. 
The passive particles are continuously driven along the circular rim of the trap, inducing a shear-like circular fluid flow inside.
In a very loose analogy, this geometry is similar to setups of Taylor-Couette flow,\cite{taylor1923stability} but, of course, here in the limit of low Reynolds numbers.
In fact, driving passive colloidal particles along ring-like trajectories can be realized experimentally via optical trapping potentials.\cite{williams2016transmission}

Considering the driven particles (that hydrodynamically interact with the interior microswimmers) as one component of a binary mixture naturally induces fluid flows to which the enclosed microswimmers are exposed.
This avoids explicitly imposing such flows as an external flow field.\cite{rauscher2007dynamic,scacchi2017dynamical,scacchi2018flow,stopper2018non}
However, we do not account in the present work for, e.g., possible effects of shear banding, which have been addressed in the context of DDFT as well.\cite{scacchi2017dynamical,scacchi2018flow,stopper2018non}
Our one-body density, particularly for passive particles within the cell, remains basically unchanged by the translational effects of the shear flow, as expected in the limits of our current theory regarding shear.\cite{brader2011density,aerov2014driven} 
Instead, for active microswimmers within the cell, the induced rotation of the swimmer orientations, coupling to the directions of self-propulsion, can lead to changes in the spatial density.

In the context of our theory, the active microswimmers represent the first species~A, while the driven colloidal particles are treated as species~B.
Consequently, $f_\mathrm{B}=0$, but we also define an effective potential of confinement
\begin{align}
u^\mathrm{B}_\mathrm{ext}(\vect{r})=&\,V^\mathrm{B}_0\Bigg(\erf\left(\frac{r-R_0-\frac{1}{2}\sigma_\mathrm{R}}{\sigma_\mathrm{R}}\right) \notag \\ &{}
- \erf\left(\frac{r-R_0+\frac{1}{2}\sigma_\mathrm{R}}{\sigma_\mathrm{R}}\right)\Bigg)
\end{align}
for the passive particles, based on the error function $\erf(s)= (2/\sqrt{\pi})\int_0^s \mathrm{d} u \exp(-u^2)$.
For $V^\mathrm{B}_0\gg k_\mathrm{B} T$ and $R_0 \gg \sigma_\mathrm{R}$, this potential effectively anchors the particles on a (small-width) ring of radius $R_0$.
Additionally, the non-conservative driving force
\begin{equation}
 \vect{F}_\mathrm{d}(\vect{r})  =\omega_\mathrm{d} \, \frac{\uvec{z}\times\vect{r}}{\mu^\mathrm{t,\mathrm{B}}}
 \label{driving_force}
\end{equation}
is taken into account to describe the continuous circular driving of the passive particles.
Technically, we include it by adding  $-\vect{F}_\mathrm{d}(\vect{r})$ to $\nabla_{\vect{r}} u^\mathrm{B}_\mathrm{ext}(\vect{r})$ in the corresponding equations.
Here, $\omega_\mathrm{d}$ is the (signed) magnitude of the imposed (spatial) angular velocity with which the passive particles are driven along the ring.

\begin{figure*}[t]
\vspace{-10pt}
 \includegraphics{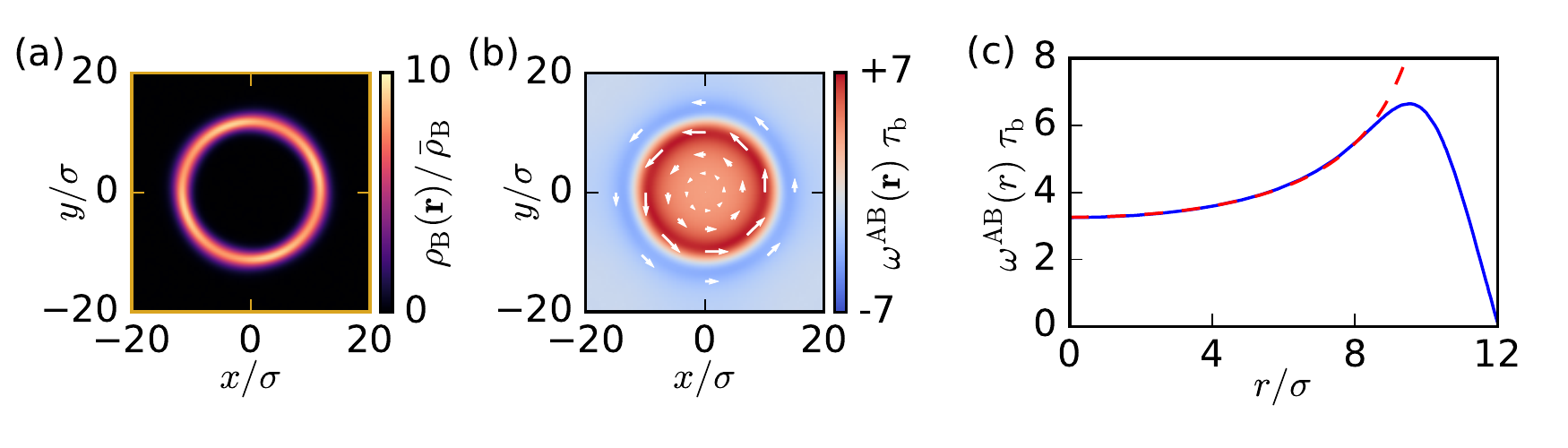}
\vspace{-20pt}
 \caption{Density ring of driven passive particles (species~B) that impose a flow field on confined microswimmers (species A) on the inside (the density of the latter not explicitly shown here). 
The system parameters are~$a_\mathrm{A}=a_\mathrm{B}=0.25 \, \sigma_\mathrm{B}$, $\epsilon^\mathrm{BB}_0=10 \, k_\mathrm{B} T$, $R_0=11.5\, \sigma_\mathrm{B}$, $\sigma_\mathrm{R}=3 \, \sigma_\mathrm{B}$, $V_0^\mathrm{B}=50 \, k_\mathrm{B} T$,  $N_\mathrm{B}=10$, 
and~$\omega_\mathrm{d} \tau_\mathrm{b}=20$ [with Brownian time $\tau_\mathrm{b}=\sigma_\mathrm{B}^2 / (\mu^{\mathrm{t},\mathrm{B}} k_\mathrm{B} T)$ and~$\sigma_\mathrm{A}=\sigma_\mathrm{B}\equiv\sigma$].
Numerically, the evaluation is performed on a $256\times256$ grid (in~$x$ and~$y$), for a simulation box of size~$20 \sigma \times 20 \sigma$. 
(a)~Ring-like density distribution of species~B (passive particles), reduced by the average density~$\bar{\rho}_\mathrm{B}$. Brighter colors represent higher densities. 
(b)~Illustration of the resulting steady hydrodynamic flows exerted on species~A (microswimmers) by species~B. 
White arrows indicate the magnitude and direction of $\vect{v}^\mathrm{AB}(\vect{r})$, according to Eq.~(\ref{vab}). 
The color code quantifies $\omega^\mathrm{AB}(\vect{r})$, according to Eq.~(\ref{wab}). 
(c)~Radial distribution of $\omega^\mathrm{AB}(r)$,
 as extracted from the full numerical evaluation [blue line, same data as in (b)] and via the semi-analytical approximation (red dashed line) given in Eq.~(\ref{wab_approx}). 
 }
\label{fig:shear_field}
\end{figure*}

For species~A, we again choose the external trapping potential defined in Eq.~(\ref{trap}), but take care when adjusting the potential strength that (even with $f_\mathrm{A} \neq 0$) it at all times hinders the majority of the swimmers from reaching the passive particles on the outer ring.
This way, species~A and B mainly interact with each other hydrodynamically, as described by, e.g., the current densities in Eqs.~(\ref{eqK1r}) and~(\ref{eqK4r}).

The driven ring of passive colloidal particles of species~B is shown in Fig.~\ref{fig:shear_field}.
For typical parameters, (a) the corresponding density profile and
(b) the hydrodynamic influences on the microswimmers of species~A are depicted.
For the latter, we define for species~A the contribution to the velocity resulting from the fluid flows induced by species~B as
\begin{equation}
 \vect{v}^\mathrm{AB}(\vect{r},t)=
 \frac{\currk{\mathrm{AB}}^{\mathrm{tt}}(\vect{X},t)}{\rho_\mathrm{A}(\vect{X},t)}
\label{vab}
\end{equation}
and the corresponding contribution to the $z$-component of the angular velocity as
\begin{equation}
 \omega^\mathrm{AB}(\vect{r},t)=\uvec{z} \cdot 
 \frac{\currk{\mathrm{AB}}^{\mathrm{rt}}(\vect{X},t)}{\rho_\mathrm{A}(\vect{X},t)} \, \,.
\label{wab}
\end{equation}
Here, the current densities, as defined in Eqs.~(\ref{eqK1r}) and~(\ref{eqK4r}) in combination with Eqs.~(\ref{eq:j_nu}) and~(\ref{Ons11}), are  proportional to $\rho_\mathrm{A}(\vect{X},t)$ so that the above expressions do not diverge when the denominator vanishes.

\begin{figure}
 \includegraphics[trim=0 50 0 45,clip]{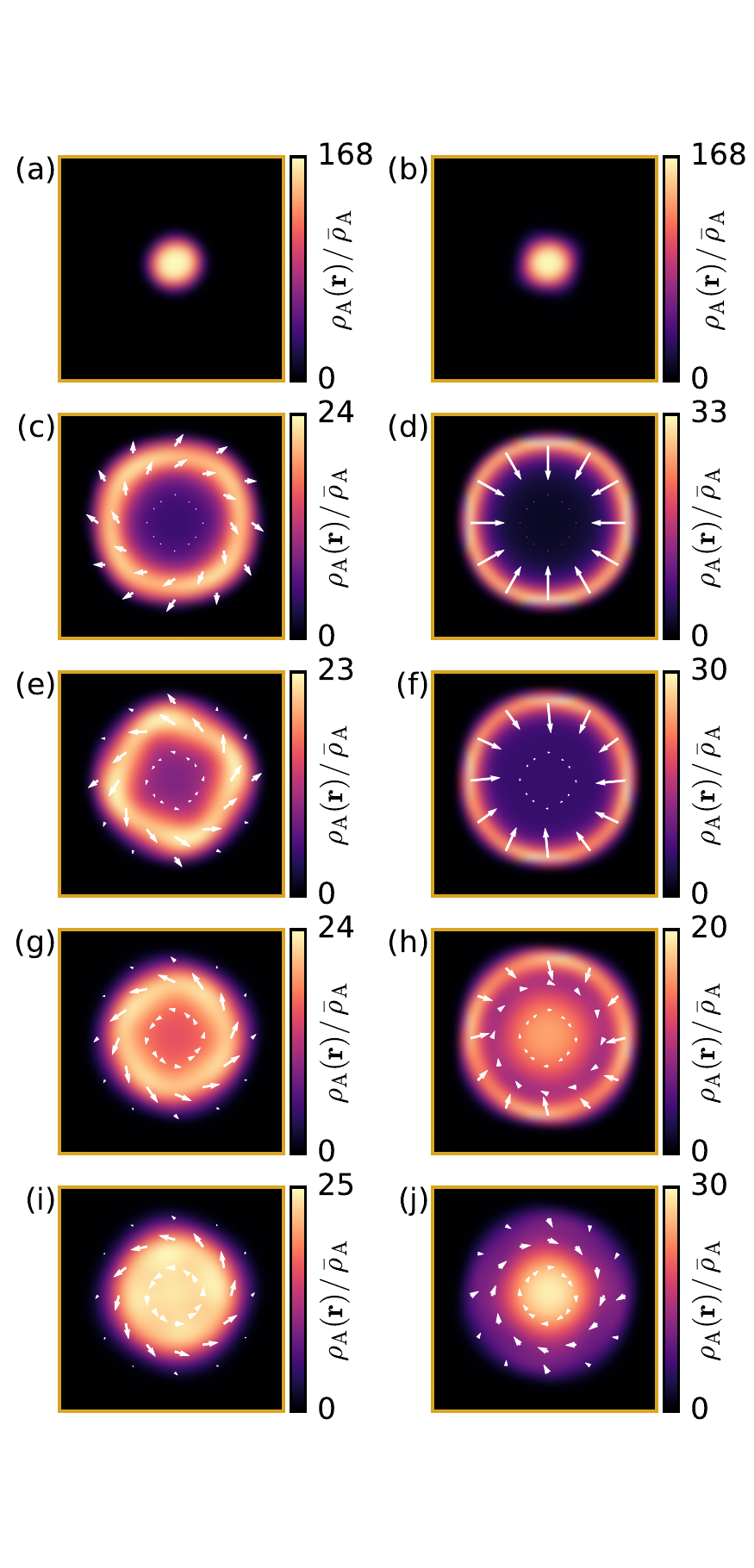}
\caption{Steady-state density distributions of species~A inside the externally driven ring of passive particles (species~B, not shown here).
In addition to the parameters (for species~B) given in Fig.~\ref{fig:shear_field}, we have used~$\bar{\rho}_\mathrm{A} \sigma^2=0.00188$, $a_\mathrm{A}=0.25 \, \sigma$, $L_\mathrm{A}=0.75 \, \sigma$, $\alpha_\mathrm{A}=0.4$, $V_0^\mathrm{A}=0.1 \, k_\mathrm{B} T$,
$\epsilon^\mathrm{AA}_0=\epsilon^\mathrm{AB}_0=10 \, k_\mathrm{B} T$,
 with $\sigma_\mathrm{A}=\sigma_\mathrm{B}=\sigma$, and only the inner area of $16 \, \sigma \times 16 \, \sigma$ is shown. 
Again, brighter colors indicate higher spatial densities and white arrows reflect the average orientation vector fields, as defined in Eqs.~(\ref{dens_profile}) and~(\ref{or_vf}), respectively.
(a,~b)~Densities of internally confined passive particles ($f_\mathrm{A}=0$), at magnitudes of the external driving (a)~$\omega_\mathrm{d}=0$ and (b)~$\omega_\mathrm{d} \tau_\mathrm{b}=80$ [with Brownian time $\tau_\mathrm{b}=\sigma^2 / (\mu^{\mathrm{t},\mathrm{B}} k_\mathrm{B} T)$]. Within the precision of our numerical discretization scheme, the distributions are identical.
(c--j)~Confined active microswimmers ($|f_\mathrm{A}|=400 k_\mathrm{B} T / \sigma$) subject to external driving strengths acting on the outer particles (c, d)~$\omega_\mathrm{d} \tau_\mathrm{b}=0$, (e, f)~$\omega_\mathrm{d} \tau_\mathrm{b}=40$, (g, h)~$\omega_\mathrm{d} \tau_\mathrm{b}=80$, and (i, j)~$\omega_\mathrm{d} \tau_\mathrm{b}=120$. Here, the cases of pushers are depicted on the left-hand side, while those for pullers are shown on the right-hand side. The induced shear flows lead to an increased localization towards the center of the cell, together with an induced tilting of the swimmer orientation, which is more pronounced for pushers than for pullers.
}
\label{fig:shear_results}
\end{figure}

The resulting density distribution of species~B depicted in Fig.~\ref{fig:shear_field} is basically circularly symmetric and after initial equilibration does not vary over time any longer.
Still, it represents the moving passive particles driven by the \mbox{(tangential)} external force defined in Eq.~(\ref{driving_force}). 
The latter is the main source of the fluid flows induced by particles of species~B. 
Resulting flow fields can be approximated inside the cell by evaluating the corresponding terms in Eqs.~(\ref{vab}) and (\ref{wab}) under the assumption of $\rho_\mathrm{B}(\vect{X}',t)\equiv N_\mathrm{B} (2 \pi)^{-2} R_0^{-1} \delta(r'-R_0)$.
Considering the contribution of~$\vect{F}_\mathrm{d}(\vect{r}')$ as dominant, ignoring steric interactions between species~A and~B, and introducing $b=r/R_0<1$, we obtain from Eq.~(\ref{wab})
\begin{align}
	\omega^\mathrm{AB}(\vect{r})\approx&\frac{3}{4} \, \frac{a}{R_0} \,  \omega_\mathrm{d} N_\mathrm{B} \, \frac{1}{2 \pi} \! \int\limits_{-\pi}^\pi \! \! \mathrm{d} \psi \, \frac{1-b \cos\psi}{\left(1 -2 b \cos \psi + b^2\right)^{3/2}} \notag \\
\approx&\frac{3}{4} \, \frac{a}{R_0} \,  \omega_\mathrm{d} N_\mathrm{B} \left(1 + \frac{3}{4} \, b^2 + \frac{45}{64} \, b^4 + \mathcal{O}(b^6) \right)
\label{wab_approx}
\end{align}
for the angular velocity.
As shown in Fig.~\ref{fig:shear_field}~(c), there is good quantitative agreement between this approximation [the integral expression in Eq.~(\ref{wab_approx}) is plotted as the dashed line]  and the full numerical solution (solid line). 
For positions close to the outer ring of the driven particles of species~B, the curve drops, most likely because of the decreased probability of finding the swimmers and the driven particles within close distances from each other, formally introduced by the Onsager-like terms in Eqs.~(\ref{Ons20})--(\ref{Ons02}). 
To leading order in $a$, the flow field induced by the driven species~B can be similarly obtained as
\begin{align}
 \vect{v}^\mathrm{AB}(\vect{r}) \approx&\frac{3}{4} \, a \, \omega_\mathrm{d} N_\mathrm{B} \left(\uvec{z} \times \uvec{r}\right) \frac{1}{\pi} \! \int\limits_{-\pi}^\pi \! \! \mathrm{d} \psi \, \frac{\cos\psi}{\left(b^2-2 b \cos \psi + 1\right)^{1/2}} \notag \\
\approx&\frac{3}{4} \, a \, \omega_\mathrm{d} N_\mathrm{B} \left(\uvec{z} \times \uvec{r}\right) \left(b + \frac{3}{8} \, b^3 + \frac{15}{64} \, b^5 + \mathcal{O}(b^7) \right).
\end{align}

\begin{figure}
 \includegraphics[trim=0 0 0 10,clip]{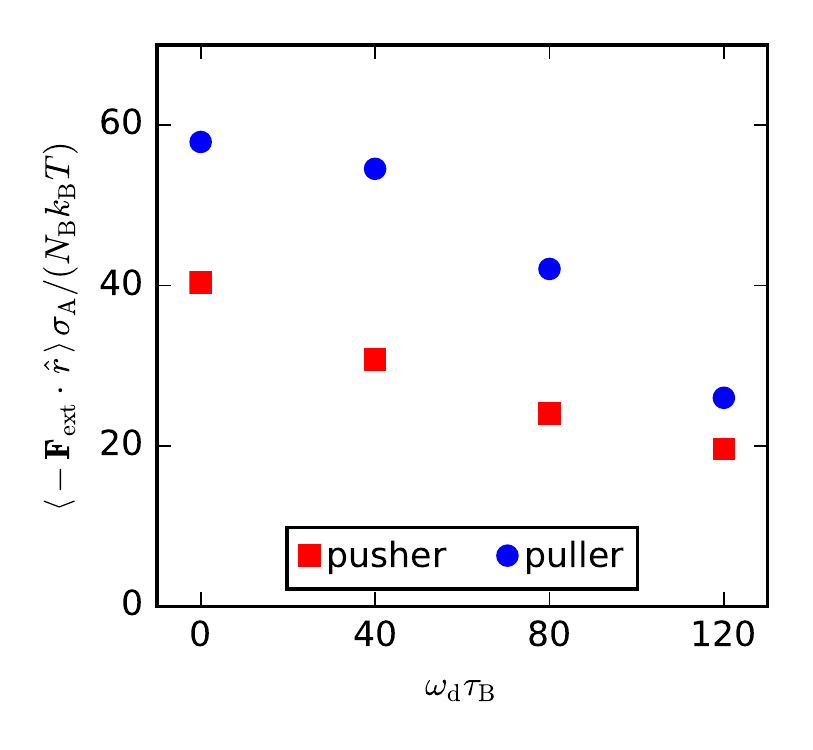}
\caption{Averaged radial component of the external force acting on the trapped microswimmers versus angular driving speed $\omega_\mathrm{d}$ of the outer passive particles, for pushers (red squares) and pullers (blue circles), resulting from the steady-state density distributions displayed in Fig.~\ref{fig:shear_results}~(c)--(j). Here, $\uvec{r}=\vec{r}/|\vec{r}|$ is the spatial unit vector pointing radially outward. With increasing $\omega_\mathrm{d}$, the swimmer orientations are rotated by the induced flow, which hinders the outward self-propulsion. This leads to increasingly centered density distributions, reducing the exposure to the external trapping in magnitude.}
\label{fig:force}
\end{figure}

We now concentrate on species~A that is confined inside the shear cell.
For $f_\mathrm{A}=0$, passive particles are recovered.
As seen in the steady states shown in Fig.~4~(a) and~(b), the distribution of the inner passive particles remains virtually unaffected by the external driving of the outer passive particles, except for possible small deviations that cannot be resolved within the precision of our numerical discretization scheme.
But when the enclosed swimmers are active ($f_\mathrm{A}\neq0$), see Fig.~\ref{fig:shear_results}~(c)--(j), the effects of the induced shear flows become significant.
Figure~\ref{fig:shear_results}~(c) and~(d) show the situation of the enclosed swimmers for pushers and pullers without the external drive, i.e., $\omega_\mathrm{d}=0$. 
Here, for the chosen parameters, the microswimmers form high-density rings with average orientations tilted relatively to the outward direction for pushers~(c) \cite{hoell2017dynamical} and radially oriented for pullers~(d).
The directional sense of the tilt for pushers is spontaneously chosen by the system as either counterclockwise or clockwise (depicted here), depending on the initialization of our numerical evaluation.
In contrast to these cases of vanishing external driving of species~B, Fig.~\ref{fig:shear_results}~(e)--(j) demonstrate that for $\omega_\mathrm{d}\neq0$, the externally induced shear flows can lead to a collapse of the steady-state density distributions towards the center of the confinement.
Moreover, with increasing external driving~$\omega_\mathrm{d}$, both pushers and pullers furthermore show an increasing tendency of their locally averaged swimming direction to be reoriented by the externally imposed fluid flow [see Fig.~\ref{fig:shear_field}~(b)].
This explains the different sense indicated by the white arrows for increased~$\omega_\mathrm{d}$ from Fig.~\ref{fig:shear_results}~(c) to Fig.~\ref{fig:shear_results}~(e).

As a source of this behavior, the shear flow induced by the external driving of the outer ring persistently rotates the orientations of the internal swimmers so that the latter are hindered from  efficiently swimming radially outwards against the trapping force.
In this way, the behavior of species~A becomes comparable to that of circle swimmers, i.e., self-propelled particles that additionally feature an active self-rotation.\cite{van2008dynamics,fily2012cooperative,kummel2013circular,lowen2016chirality,sevilla2016diffusion} 
Actually, we have observed a similar phenomenology as in Fig.~\ref{fig:shear_results} for increasing inherent curvature of the trajectories of circle swimmers in Ref.~\onlinecite{hoell2017dynamical}.
In the present case, however, the (externally induced) rotation varies with the distance $r$ from the origin, so that the local radius of induced circle-swimming $R_\mathrm{cs}(r):=|\vect{v}_{0\mathrm{A}} / \omega_\mathrm{AB}(r)|$, determined from the definitions in Eqs.~(\ref{v0}) and~(\ref{wab_approx}), is non-constant.
It reaches a maximum at the origin and decreases with increasing~$r$.
For Fig.~\ref{fig:shear_results}~(e)--(h), the length scale of~$R_\mathrm{cs}(r)$ is comparable to the radius of the effective trap so that a high-density ring is still visible.
However, the average orientations are significantly tilted from the radial direction (especially for pusher microswimmers).
Further increasing the external driving strength, see Fig.~\ref{fig:shear_results}~(i) and~(j), leads to more localized density profiles and circling around the center of confinement.

The increasing localization can be quantified by the (negative) radial component of the averaged external trapping force experienced by the microswimmer ensemble, as given in Fig.~\ref{fig:force} for the same (steady-state) data as in Fig.~\ref{fig:shear_results}~(c)--(j). 
For vanishing angular driving speed $\omega_\mathrm{d}$ of the outer passive particles, we find a higher value for pullers (blue circles) than for pushers (red squares), corresponding to the more off-center density distribution of pullers caused by their stronger tendency to show radial orientation.
Both curves drop for increasing $\omega_\mathrm{d}$. 
The reason is again the induced shear flow increasingly hindering the swimmers from self-propelling efficiently against the radial external trapping potential.
The drop is somewhat delayed for our pullers, in accordance with a similar effect previously seen for circle swimmers in an external trap, where the pullers also showed a stronger tendency of maintaining a ring of outward-oriented swimmers.\cite{hoell2017dynamical}

In related works, rosette-like trajectories have been reported for (single) circle swimmers with explicitly time-dependent self-propulsion velocities.\cite{babel2014swimming,jahanshahi2017brownian}
Beyond the scope of the present work, when genuine circle swimmers are confined (as species~A) in our setup, again high-density rings might be observed with average swimmer orientations along the local radial direction.
For this purpose, the induced rotation $\omega^\mathrm{AB}$ should balance the inherent self-rotation of the circle swimmers.

\section{Conclusions}
\label{sec:con}

In this work, we have presented a dynamical density functional theory (DDFT) for multi-species suspensions of microswimmers.
We have included (pairwise) hydrodynamic and effective steric interactions between swimmers.
In this way, we conceptually extended the previous one-component equivalent.\cite{menzel2016dynamical,hoell2017dynamical,hoell2018particle}
The theory is based on a discrete force-dipole minimal microswimmer model, which has already been used successfully in several previous works.\cite{menzel2016dynamical,hoell2017dynamical,pessot2018,hoell2018particle,daddi2018dynamics}
We then applied our theory to three illustrative example situations of planar swimmer configurations inside a three-dimensional bulk fluid.

First, binary pusher--puller mixtures in external spherically symmetric trapping potentials have been discussed.
For the two species only differing in their pusher~/~puller signature, we found that the majority species imposes its behavior on the minority species. 
For example, pushers at the considered propulsion strength on their own tend to form concentrated spots on the rim of the trap. 
Therefore, if pushers represent the majority species, this spot formation is conveyed to simultaneously present pullers. 
Conversely, pullers by themselves rather tend to form a roughly spherically symmetric high-density ring on the rim of the trap. 
Thus, if they represent the majority in a pusher--puller mixture, also pushers tend to organize themselves in a corresponding ring structure.

Second, in the absence of any external trapping potential, pusher--puller mixtures in large periodic boxes have been considered. 
In an analytical treatment analogous to the previously studied one-component case,\cite{hoell2018particle} pullers are found to be able to establish the onset of collective polar orientational order of the whole mixture.
Accordingly, pullers can induce oriented collective motion.
For this purpose, they need to represent the majority species and show a sufficiently large magnitude of their active drive. 
Our results are qualitatively in line with previous agent-based computer simulations.\cite{pessot2018}

Third, a microswimmer species is confined inside a circular ring of externally driven passive particles. 
The induced shear flow persistently rotates the confined swimmers, and thus can hinder them from forming the high-density rings that are typically observed for sufficiently quick self-propelled particles in radial trapping potentials. 
Instead, the swimmer densities tend to collapse towards the center of the confinement.
Similar mechanisms have previously been found for circle swimmers (featuring an inherent self-rotation) without externally induced shear flows.
One future task could be to focus further on the role of shear flows in our statistical theory.\cite{brader2011density,aerov2014driven}

In the numerical examples, we have restricted our evaluations for hydrodynamically interacting microswimmers to small confined systems that suitably fit into the corresponding simulation box. 
Nevertheless, in the future, our set of partial differential equations could be solved numerically as well for (basically infinitely extended) bulk situations. 
For this purpose, (true) periodic boundary conditions are applied to a finite simulation box.
Then, because of the long-range nature of the hydrodynamic interactions, the influence of all periodic images on the density distribution in the simulation box must be accounted for. 
Mathematically, this can be achieved by applying Ewald summation techniques \cite{ewald1921berechnung} to the mobility tensors.
Corresponding results have been derived for passive particles,\cite{beenakker1986ewald,brady1988dynamic,rex2008influence} but could,
in principle, also be calculated for our active microswimmers, 
as has recently been demonstrated for a similar force-dipole-based microswimmer model.\cite{adhyapak2018ewald}
For quantitative tests of our theory in the future and for extensions to higher densities, particle-based computer simulations (that include hydrodynamic flows of the surrounding fluid and thermal fluctuations explicitly) can be performed.

One very interesting question is whether our DDFT could be extended to describe the aforementioned motility-induced phase separation. 
In this context, existing statistical theories involved a density-dependent effective swimming speed and/or an anisotropic pair distribution function as additional inputs.\cite{cates2013active, bialke2013microscopic, speck2014effective, speck2015dynamical,wittkowski2017nonequilibrium} 
It would thus be interesting to study in the future the effect of at least one similar activity-induced term in our theory.
Another promising statistical approach beyond the adiabatic approximation of DDFT is the power functional theory for ``dry'' self-propelled particles, which has recently been formulated and evaluated semi-analytically.\cite{krinninger2016nonequilibrium,krinninger2017erratum,krinninger2019power}

The present work derives the multi-species DDFT for the case of uniaxial straight-swimming microswimmers with spherical bodies.
However, only a few changes transfer it to the case of (inherently biaxial) circle swimmers.\cite{hoell2017dynamical}
Even more generally, changes will allow to describe swimmers with less-symmetric body shapes, e.g., rod-like, bodies.
Nevertheless, we remark that more work is needed in the future regarding situations of still higher complexity.
Examples are cases in which, for instance, additional phoretic chemical- or temperature-based interactions between swimmers become significant.\cite{liebchen2019interactions}
Moreover, effects of the fluctuations of the propulsion mechanism itself could be taken into account.\cite{pietzonka2017entropy}

Beyond the direct numerical evaluations performed in this work, DDFTs can serve as a foundation to derive corresponding phase-field-crystal models\cite{van2009derivation,menzel2013traveling,menzel2014active,wittkowski2014scalar,alaimo2018microscopic} and more macroscopic continuum theories\cite{heidenreich2016hydrodynamic,reinken2018derivation,reinken2019anisotropic} of microswimmer suspensions.
The latter allow for connections towards still-larger length scales of theoretical descriptions.
Altogether, we thus expect our DDFT to provide a powerful tool for the statistical characterization of dynamic multi-species systems of suspended microswimmers, of future relevance both in fundamental physics and concerning the corresponding biological, technical, and medical applications.

\begin{acknowledgments}
The authors thank Urs Zimmermann, Soudeh Jahanshahi, and Giorgio Pessot for helpful comments, as well as the Deutsche Forschungsgemeinschaft for support of this work through the SPP 1726 on microswimmers, Grant Nos.\ LO 418/17-2 and ME 3571/2-2.
\end{acknowledgments}

\bibliographystyle{apsrev}
\bibliography{ddft-mixture}

\end{document}